\documentclass[12pt]{amsart}

\usepackage[hmargin=0.8in,twosideshift=0in,height=8.6in]{geometry}
\usepackage{amssymb,amsthm, times}
\usepackage{delarray,verbatim}
\usepackage{natbib}
\usepackage{bm}
\usepackage{ifpdf}
\ifpdf
\usepackage[pdftex]{graphicx}
 \else
\usepackage[dvips]{graphicx}
 \fi

\usepackage{ifpdf}
\usepackage{color}
\definecolor{webgreen}{rgb}{0,.5,0}
\definecolor{webbrown}{rgb}{.8,0,0}
\definecolor{emphcolor}{rgb}{0.95,0.95,0.95}

\usepackage{hyperref}
\hypersetup{%
          colorlinks=true,
          linkcolor=webbrown,
          filecolor=webbrown,
          citecolor=webgreen,
          breaklinks=true}
\ifpdf \hypersetup{pdftex,
            pdfstartview=FitH, %%Fit, FitB, FitH
            bookmarksopen=true,
            bookmarksnumbered=true
} \else \hypersetup{dvips} \fi

\numberwithin{equation}{section}

\newtheorem{theorem}{Theorem}[section]

\newtheorem{proposition}{Proposition}[section]
\newtheorem{corollary}{Corollary}[section]
\newtheorem{remark}{Remark}[section]
\newtheorem{lemma}{Lemma}[section]
\newtheorem{example}{Example}[section]

\newtheorem{definition}{Definition}[section]
\numberwithin{remark}{section} \numberwithin{proposition}{section}
\numberwithin{corollary}{section}
\newcommand {\R}{\mathbb{R}}

\newcommand {\p}{\mathbb{P}}

\newcommand {\E}{\mathbb{E}}

\newcommand{\diff}{{\rm d}}

\newcommand{\lev}{L\'{e}vy }

\title[solving optimal dividend problems via phase-type fitting approximation of scale functions]{solving optimal dividend problems via phase-type fitting approximation of scale functions}
\author[M. Egami]{Masahiko Egami}
\address[M. Egami]{Graduate School of Economics,
Kyoto University, Sakyo-Ku, Kyoto, 606-8501, Japan }
\email{egami@econ.kyoto-u.ac.jp}
\thanks{M. Egami is in part supported
by Grant-in-Aid for Scientific Research (C) No. 20530340, Japan
Society for the Promotion of Science. K. Yamazaki is supported by Grant-in-Aid for Young Scientists (B) No. 22710143, the Ministry of Education, Culture, Sports, Science and
Technology.}
\author[K. Yamazaki]{Kazutoshi Yamazaki }
\address[K. Yamazaki]{Center for the Study of Finance and Insurance,
Osaka University, 1-3 Machikaneyama-cho, Toyonaka City, Osaka 560-8531, Japan}
\email{k-yamazaki@sigmath.es.osaka-u.ac.jp}
\thanks{}

\begin{document}

\begin{abstract}
\noindent The optimal
dividend problem by \cite{Definetti_1957} has
been recently generalized to the spectrally negative \lev model
where the implementation of optimal strategies draws upon the
computation of scale functions and their derivatives. This paper
proposes a phase-type fitting approximation of the optimal strategy.
We consider spectrally negative \lev processes with phase-type jumps
as well as meromorphic \lev processes \citep{Kuznetsov_2010}, and use their
scale functions to approximate the scale function for a general
spectrally negative \lev process.  We obtain analytically the
convergence results and illustrate numerically the effectiveness of
the approximation methods using examples with the spectrally negative
\lev process with i.i.d.\ Weibull-distributed jumps, the $\beta$-family and CGMY process.

\end{abstract}

\date{\today}

\maketitle
\noindent \small{\textbf{Key words:}
De Finetti's dividend problem; phase-type models; Meromorphic \lev processes; Spectrally negative \lev processes; Scale functions 
%\noindent \small{\textbf{JEL Classification:}\, G13,  G33, D81, C61 }
 }\\

\section{Introduction}
In the optimal dividend problem, an insurance company wants to
maximize the cumulative amount of dividends paid out to its
beneficiaries until the time of ruin. The original problem by
\cite{Definetti_1957} considered a discrete-time case by modeling
the surplus by a random walk.  It has been later extended to the
continuous-time  diffusion model
\citep{Shiryaev_1995,Asmussen_1997,Gerber_Shiu_2004} and to the
Cram\'er-Lundberg model \citep{Gerber_1969, Azcue_Muler_2005}. It
was recently generalized to the \emph{spectrally negative \lev
model} by \cite{Avram_et_al_2007} and
\cite{Kyprianou_Palmowski_2007} where the surplus is a general \lev
process with only negative jumps.

The implementation of the optimal strategies in the spectrally
negative \lev model draws upon the computation of the so-called
\emph{scale function}.  \cite{Avram_et_al_2007} first obtained the
expected value under a barrier strategy in terms of the scale
function. \cite{Loeffen_2008} then showed that a barrier strategy is
indeed optimal under a suitable condition.  The scale function also
plays great roles in its extensions with transaction costs and
additional terminal values at ruin. See \cite{Loeffen_2009,
Loeffen_2009_2} and \cite{Loeffen_2010}.

Despite these advances, a major obstacle still remains in putting in
practice the above-mentioned results because scale functions are in
general known only up to their Laplace transforms. The
implementation is even harder for the optimal dividend problem
because it requires the first derivative of the scale function. One
can in principle approximate the scale function and its derivative
via the numerical Laplace inversion as discussed in
\cite{Surya_2008}.  However, the method is only heuristic and one
cannot determine the accuracy of the approximation.  Moreover, the
derivative of the scale function tends to explode in the
neighborhood of zero, and therefore the error is expected to be
large near zero. For these reasons, there is a clear need of new
approximation methods that work for any spectrally negative \lev
process.

As a new tool to overcome these problems, this paper proposes a
\emph{phase-type fitting} approach to approximate the scale
functions as well as the solutions to the optimal dividend problem and its extensions.
We obtain the scale functions of spectrally negative \lev processes
with phase-type jumps and those in the M-class (meromorphic \lev
processes), and apply these for a general spectrally negative \lev model.

We first consider the class of spectrally negative \lev processes
with phase-type jumps. Consider a continuous-time Markov chain with
some initial distribution and state space consisting of a single
absorbing state and a finite number of transient states.  The
phase-type distribution is the distribution of the time to
absorption.
%The class of phase-type distributions includes, for example, the exponential, hyperexponential,  Erlang and Coxian distributions; see Section 3 of Asmussen \cite{Asmussen_2003}.
It is known that the class of phase-type distributions is \emph{dense} in the class of all positive-valued distributions. We obtain the scale functions for these processes and show that they can approximate the scale function of a general spectrally negative \lev process arbitrarily closely.

The phase-type fitting approach has mainly three advantages. First,
thanks to the smoothness and monotonicity properties of scale functions as proved
by, for example, \cite{Chan_2009} and \cite{Loeffen_2008}, the approximation can
be applied also to its derivative.
%It should be noted that the
%derivative is used to identify the fluctuation of its reflected
%process and is commonly applied in insurance risk literature (Avram
%et al.\ \cite{Avram_et_al_2007}, Kyprianou and Palmowski
%\cite{Kyprianou_Palmowski_2007} and Kyprianou and Zhou
%\cite{Kyprianou_Zhou_2009}).
Second, the Laplace transform of the
phase-type distribution has an explicit expression and hence can avoid the error caused while approximating the
Laplace transform for a general jump distribution. This type of
errors tends to occur in other approximation methods such as
\cite{Surya_2008}.
%\cite{Feldmann_1998}
%address that one of their motivations of approximating via
%hyperexponential distributions comes from the analytical
%tractability of obtaining their Laplace transforms.
Third, the phase-type fitting approach enjoys a
variety of fitting algorithms. See, for example, \cite{ Asmussen_1996},
\cite{Bladt_2003} and
\cite{Feldmann_1998}. The fitting can be applied also to empirical
data and this is another major advantage.  In the first half of our numerical results, we use the results by \cite{Feldmann_1998} and consider the case with a Brownian motion plus a compound Poisson with Weibull-distributed jumps.

We also consider meromorphic \lev processes \citep{Kuznetsov_2010}, which generalizes a number of recently-discovered \lev processes such as Lamperti-stable processes \citep{Caballero_2006, Chaumont_2009},  hypergeometric processes \citep{Kyprianou_2010, Kuznetsov_2010_3} and processes in the $\beta$- and $\Theta$-families \citep{Kuznetsov_2010_2,Kuznetsov_2009}. The spectrally negative versions of these processes commonly have \lev measures in the form
\begin{align}
\nu(\diff z) =  \left[ \sum_{i=1}^\infty \alpha_i \eta_i e^{- \eta_i
|z|} 1_{\{z < 0\}} \right] \diff z,  \quad z
\in \mathbb{R}, \label{density_meromorphic}
\end{align}
which can be seen as an extension of the hyperexponential density and as a ``discrete version" of  the completely monotone density.

The reason we consider the \lev measure above is that it can
approximate efficiently any \lev measure with a completely monotone
density. Furthermore, if the \lev measure is completely monotone,
the barrier-strategy is guaranteed to be optimal for the dividend problem
\citep{Loeffen_2008}.  The class of \lev measures with completely
monotone densities is rich. It enables us to model
compound-Poisson-type jumps with long-tailed distributions such as
the Pareto, Weibull and gamma distributions; see
\cite{Feldmann_1998}.   It further allows us to construct many of
recently-introduced \lev processes such as variance gamma processes
\citep{Madan_1991,Madan_1998}, CGMY processes \citep{CGMY_2002},
generalized hyperbolic processes \citep{Eberlein_1998} and normal
inverse Gaussian processes \citep{Barndorff_1998}.  These processes
can be approximated efficiently by meromorphic \lev processes.

Given a \lev measure in the form (\ref{density_meromorphic}), the
corresponding scale function and its derivative can be expressed
explicitly as \emph{infinite} sums of exponential functions.  We
first show  that these can be approximated by \emph{finite} sums with some
analytical error bounds, concluding that the error
bounds for the solutions to the optimal dividend problem can also be
obtained. We then show numerically the effectiveness of the
approximation procedure using examples with $\beta$-processes
\citep{Kuznetsov_2010_2} and CGMY processes. We obtain bounds on
scale functions and solutions for the former and use them to
approximate for the latter.

It should be emphasized here that the approximation procedure
discussed in this paper can be applied outside the class of optimal
dividend problems. The spectrally negative \lev model has been
recently introduced widely and the scale function plays a great role
in characterizing the solutions.  We refer the reader to
\cite{Avram_2004} and \cite{Alili2005} for derivative pricing,
\cite{Kyprianou_Surya_2007} for optimal capital structure,
\cite{Baurdoux2008,Baurdoux2009}  for stochastic games.  For a
comprehensive account, see \cite{Kyprianou_2006}.

The rest of the paper is organized as follows.  Section \ref{section_dividend_problem} summarizes the results on the classical dividend problem and its extensions.  In Section  \ref{section_phase_type}, we obtain the scale functions for spectrally negative \lev processes with phase-type jumps and show that it can approximate the scale function of a general spectrally negative \lev process arbitrarily closely.  We then obtain in Section \ref{section_meromorphic} the scale functions for those in the M-class including its upper and lower bounds.   We conclude this paper by giving numerical results in Section \ref{section_numerical_results}. All proofs are given in the appendix.

\section{Optimal Solutions to Dividend Problem via Scale Functions} \label{section_dividend_problem}
This section reviews the classical dividend problem and its extensions focusing on the spectrally negative \lev model.
Let $(\Omega,\mathcal{F},\mathbb{P})$ be a probability space hosting a \emph{spectrally negative} \lev process $X = \left\{X_t; t \geq 0 \right\}$ with its \emph{Laplace exponent}
\begin{align}
\psi(s)  := \log \E \left[ e^{s X_1} \right] =  \hat{\mu} s +\frac{1}{2}\sigma^2 s^2 + \int^0_{-\infty} (e^{s z}-1 - s z 1_{\{z > -1\}}) \nu (\diff z), \quad s \in \mathbb{C} \label{laplace_spectrally_negative}
\end{align}
where $\nu$ is a \lev measure with the support $(-\infty,0)$ that satisfies the integrability condition $\int_{(-\infty,0)} (1 \wedge z^2) \nu(\diff z) < \infty$.  Moreover, let $\mathbb{P}^x$ be the conditional probability under which $X_0 = x$ (also let $\mathbb{P} \equiv \mathbb{P}^0$), and $\mathbb{F} := \left\{ \mathcal{F}_t: t \geq 0 \right\}$ be the filtration generated by $X$.  The process $X$ is called the \emph{\lev insurance risk process} (or the \emph{risk process }in short), and models the surplus of an insurance company before dividends are deducted.

%\subsection{Dividend Problem}
The classical dividend problem is a control problem where the cumulative amount of dividends prior to ruin is maximized.
A (dividend) \emph{strategy} $\pi := \left\{ L_t^{\pi}; t \geq 0 \right\}$ is given by a \emph{nondecreasing}, \emph{left-continuous} and \emph{$\mathbb{F}$-adapted} process starting at zero.    Corresponding to every strategy $\pi$, the remaining amount of surplus after dividends are deducted is given by $U^\pi = \{U_t^\pi: t \geq 0 \}$ where
\begin{align*}
U_t^\pi := X_t - L_t^\pi, \quad t \geq 0,
\end{align*}
and its \emph{ruin time} is the first time it goes below zero:
\begin{align*}
\sigma^\pi = \inf \left\{ t > 0: U_t^\pi < 0 \right\}.
\end{align*}
A lump-sum payment must be smaller than the available fund and hence it is required that
\begin{align}
L_{t+}^\pi - L_t^\pi \leq U_t^\pi, \quad t < \sigma^\pi \; \; a.s. \label{admissibility}
\end{align}
Let $\Pi$ be the set of all admissible strategies satisfying (\ref{admissibility}).
The problem concerns the expected sum of total discounted dividends until ruin
\begin{align*}
v_\pi(x) := \E^x \left[ \int_0^{\sigma^\pi} e^{-q t} \diff L_t^\pi \right], \quad \pi \in \Pi,
\end{align*}
and wants to obtain an admissible strategy that maximizes it.  Hence the classical dividend problem is written as
\begin{equation}\label{eq:classical-p}
  v(x):=\sup_{\pi\in \Pi}v_\pi(x).
\end{equation}

\subsection{Scale functions} The solutions to the dividend problems can be written in terms of the scale function.  Here, we describe the scale function and summarize its properties that will be used in this paper.

Fix $q \geq 0$ and any spectrally negative \lev process with its Laplace exponent $\psi$ as defined in (\ref{laplace_spectrally_negative}). The scale function $W^{(q)}: \R \mapsto \R$ is a function whose Laplace transform is given by
\begin{align}\label{eq:scale}
\int_0^\infty e^{-s x} W^{(q)}(x) \diff x = \frac 1
{\psi(s)-q}, \qquad s > \zeta_q
\end{align}
where
\begin{align}
\zeta_q :=\sup\{s  \geq 0: \psi(s)=q\}, \quad
q\ge 0. \label{zeta}
\end{align}
We assume $W^{(q)}(x)=0$ on $(-\infty,0)$.

Let us define the \emph{first down-} and \emph{up-crossing times}, respectively, by
\begin{align}
\label{first_passage_time}
\tau_a := \inf \left\{ t \geq 0: X_t < a \right\} \quad \textrm{and} \quad \tau_b^+ := \inf \left\{ t \geq 0: X_t >  b \right\},
\end{align}
for every $0 \leq a \leq x \leq b$. Then we have
\begin{align}
\E^x \left[ e^{-q \tau_b^+} 1_{\left\{ \tau_b^+ < \tau_0 \right\}}\right] = \frac {W^{(q)}(x)}  {W^{(q)}(b)} \quad \textrm{and} \quad
\E^x \left[ e^{-q \tau_0} 1_{\left\{ \tau_b^+ > \tau_0 \right\}}\right] = Z^{(q)}(x) -  Z^{(q)}(b) \frac {W^{(q)}(x)}  {W^{(q)}(b)} \label{laplace_in_terms_of_z}
\end{align}
where
\begin{align}
Z^{(q)} (x) := 1 + q  \int_0^x W^{(q)} (y) \diff y, \quad x \in \R. \label{def_z}
\end{align}
Here, we disregard the case when $X$ is a negative subordinator (or decreasing a.s.).

We also consider a version of the scale function
$ W_{\zeta_q}: \R \mapsto \R$ that satisfies
\begin{align}
W^{(q)} (x) = e^{\zeta_q x} W_{\zeta_q} (x), \quad x \in \R \label{w_phi}
\end{align}
with its Laplace transform
\begin{align}
\int_0^\infty e^{-s x} W_{\zeta_q} (x) \diff x &= \frac 1 {\psi(s+\zeta_q)-q}, \quad s > 0. \label{scale_version}
\end{align}
Suppose $\mathbb{P}_c$, for any given $c > 0$, is the probability measure defined by the Esscher transform
\begin{align*}
\left. \frac {\diff \mathbb{P}_c} {\diff \mathbb{P}} \right|_{\mathcal{F}_t} = e^{c X_t - \psi(c) t}, \quad t \geq 0;
\end{align*}
see page 78 of \cite{Kyprianou_2006}.
Then $W_{\zeta_q}$ under $\mathbb{P}_{\zeta_q}$ is analogous to $W^{(0)}$ under  $\mathbb{P}$. Furthermore, it is known that $W_{\zeta_q}$ is monotonically increasing and
\begin{align}
W_{\zeta_q} (x) \nearrow  {(\psi'(\zeta_q))^{-1}} \quad \textrm{as} \; x \rightarrow \infty, \label{scale_function_asymptotic2}
\end{align}
which also implies that the scale function $W^{(q)}$ increases exponentially in $x$;
\begin{align}
W^{(q)} (x) \sim \frac {e^{\zeta_q x}} {\psi'(\zeta_q)} \quad \textrm{as } \; x \rightarrow \infty. \label{scale_function_asymptotic}
\end{align}
Due to the fact that $W_{\zeta_q}$ does not explode for large $x$ as opposed to $W^{(q)}$, it is often convenient to deal with $W_{\zeta_q}$ and convert it to $W^{(q)}$ using (\ref{w_phi}), especially when numerical computations are involved; see \cite{Surya_2008}.

Recall that  a spectrally negative \lev process has paths of bounded variation if and only if
\begin{align*}
\sigma = 0 \quad \textrm{and} \quad \int_{(-\infty,0)} (1 \wedge |x|) \nu(\diff x) < \infty;
\end{align*}
see, for example, \cite{Kyprianou_2006}, Lemma 2.12. In this case, we can rewrite the Laplace exponent (\ref{laplace_spectrally_negative}) by
\begin{align*}
\psi(s) = \mu s + \frac 1 2 \sigma^2 s^2 + \int_{-\infty}^0 (e^{sz} - 1) \nu (\diff z), \quad s \in \mathbb{C},
\end{align*}
with
\begin{align*}
\mu := \hat{\mu} - \int_{-1}^0 z \nu (\diff z).
\end{align*}

Regarding the smoothness of the scale function, it has been shown by \cite{Chan_2009} that if a \lev process has a Gaussian component ($\sigma > 0$), we have $W^{(q)} \in C^2(0,\infty)$. When it does not have a Gaussian component  and if its jump distribution has no atoms, we have $W^{(q)} \in C^1(0,\infty)$. In particular, a stronger result holds for the completely monotone jump case.  Recall that a density function $f$ is called completely monotone if all the derivatives exist and, for every $n \geq 1$,
\begin{align*}
(-1)^n f^{(n)} (x) \geq 0, \quad x \geq 0,
\end{align*}
where $f^{(n)}$ denotes the $n^{th}$ derivative of $f$.

\begin{lemma}[\cite{Loeffen_2008}] \label{lemma_completely_monotone}
If the \lev measure has a completely monotone density,
$W_{\zeta_q}$ is again completely monotone.
%;  $W_{\zeta_q} \in C^\infty (0,\infty)$ and
%\begin{align*}
%(-1)^n W_{\zeta_q}^{(n)}(x) \geq 0, \quad x \geq 0.
%\end{align*}
\end{lemma}

Finally, the behavior in the neighborhood of zero is given as follows. See Lemmas 4.3 and 4.4 of \cite{Kyprianou_Surya_2007}.
\begin{lemma} \label{lemma_zero}
For every $q \geq 0$, we have
\begin{align*}
W^{(q)} (0) = \left\{ \begin{array}{ll} 0, & \textrm{unbounded variation} \\ \frac 1 {\mu}, & \textrm{bounded variation} \end{array} \right\} \quad \textrm{and} \quad
W^{(q)'} (0+) = \left\{ \begin{array}{ll}  \frac 2 {\sigma^2}, & \sigma > 0 \\   \infty, & \sigma = 0 \; \textrm{and} \; \nu(-\infty,0) = \infty \\ \frac {q + \nu(-\infty,0)} {\mu^2}, & \textrm{compound Poisson} \end{array} \right\}.
\end{align*}
\end{lemma}

\subsection{Solutions in terms of scale functions}

In a spectrally negative \lev model, the expected value under the \emph{barrier strategy} can be expressed in terms of the scale function as shown by \cite{Avram_et_al_2007}. A barrier strategy at level $a \geq 0$ is denoted by $\pi_a := \left\{ L_t^a; t \leq \sigma_a \right\}$ where
\begin{align*}
L_t^a := \sup_{s \leq t} (X_s - a) \vee 0, \quad t \geq 0.
\end{align*}
We further let $\sigma_a := \inf \left\{ t > 0: U_t^{\pi_a} < 0 \right\}$ denote the corresponding ruin time.

%Now the value under the barrier-strategy can be expressed in terms of the scale function as in the following.
\begin{theorem}[\cite{Avram_et_al_2007}, (5.1)] \label{theorem_optimality} For every $a \geq 0$, we have
\begin{align} \label{value_original}
v_{\pi_a}(x) = \left\{ \begin{array}{ll} u_a(x), & x \leq a, \\ x-a + u_a(a), & x > a, \end{array} \right.
\end{align}
where
\begin{align}
u_a(x) :=\frac {W^{(q)}(x)} {W^{(q)'}(a)}. \label{def_u}
\end{align}
\end{theorem}
%In the theorem above, we assume that the derivative $W^{(q)'}$ exists and this is satisfies whenever the \lev measure $\nu$ does not have atoms.  Additional smoothness properties of the scale function have been proved.  In particular, if the \lev measure is completely monotone, we have $W^{(q)} \in C^\infty(0,\infty)$ and $W^{(q)'}$ is log-convex.
Furthermore, the barrier strategy attains optimality under a suitable condition.
\begin{theorem}[\cite{Loeffen_2008}] \label{optimality_barrier}
Suppose $a^*$ satisfies
\begin{align*}
W^{(q)'}(a) \leq W^{(q)'}(b), \quad a^* \leq a \leq b.
\end{align*}
Then the barrier strategy $\pi_{a^*}$ is an optimal strategy.
\end{theorem}
In view of the above, an optimal barrier strategy exists, for example, when $W'$ is convex, which holds whenever \lev measure has a completely monotone density as in Lemma \ref{lemma_completely_monotone}.

\textbf{Extension with bail-out.}
A variant called the \emph{bail-out problem}  is discussed in \cite{Avram_et_al_2007}.  Here the beneficiary of the dividends must inject capital to keep the risk process from going below zero.   A strategy  is now a pair $\overline{\pi} = \left\{ L^{\overline{\pi}}, R^{\overline{\pi}}\right\}$ where $L^{\overline{\pi}}$ is the cumulative amount of dividends as in the classical model and $R^{\overline{\pi}}$ is a right-continuous process representing the cumulative amount of injected capital satisfying
\begin{align}
\int_0^\infty e^{-qt} \diff R_t^{\overline{\pi}} < \infty, \quad a.s. \label{admissibility2}
\end{align}
Assume that $\varphi > 1$ is the cost per unit injected capital, the problem is to maximize
\begin{align*}
\overline{v}_{\overline{\pi}} (x) = \mathbb{E}^x \left[ \int_0^\infty e^{-q t} \diff L_t^{\overline{\pi}} - \varphi \int_0^\infty e^{-q t} \diff R_t^{\overline{\pi}}\right]
\end{align*}
among all strategies that satisfy (\ref{admissibility}) and (\ref{admissibility2}). In this model, the optimal strategy reduces to the \emph{double-barrier strategy} that regulates the risk process $V_t := X_t - L_t^{\overline{\pi}} + R_t^{\overline{\pi}}$ inside the interval $[0,d^*]$ where
\begin{align} \label{optimal_barrier_bail_out}
d^* := \inf \left\{a > 0: G(a) := (\varphi Z^{(q)} (a) - 1 ) W^{(q)'}(a) - \varphi q W^{(q)} (a)^2 \leq 0 \right\}.
\end{align}
The value function becomes $\overline{v}_{d^*}(x)$ with
\begin{align}
\overline{v}_d(x) = \left\{ \begin{array}{ll} \varphi \left( \int_0^x Z^{(q)}(y) \diff y + \frac {\psi'(0+)} q \right) + Z^{(q)}(x) \left[ \frac {1-\varphi Z^{(q)}(d)} {q W^{(q)}(d)}\right], & 0 \leq x \leq d, \\ x - d + \overline{v}_d(d), & x > d.\end{array} \right. \label{def_v_bar}
\end{align}
%For more details, see \cite{Avram_et_al_2007}.

\textbf{Extension with terminal values at ruin.}
\cite{Loeffen_2010} recently considered the case with additional terminal costs at ruin; the objective function is
\begin{align*}
\max_{\pi \in \Pi} \E^x \left[ \int_0^{\sigma^\pi} e^{-q t} \diff L_t^\pi  + e^{-q \sigma^\pi} P(U^\pi_{\sigma^\pi}) 1_{\{ \sigma^\pi < \infty \}}\right]
\end{align*}
for some affine function $P(y) := S + K y$.  Its special case with constant terminal value was studied by \cite{Loeffen_2009}. With the assumption that the tail of the \lev measure is log-convex, the optimal solution is either the barrier strategy or the \emph{take-the-money-and-run strategy} where the latter immediately pays out all the dividends and forces the ruin to occur immediately. For the former case, the optimal barrier level is given by
\begin{align*}
b^* := \sup \left\{ b \geq 0: F(b) \geq F(x) \; \textrm{for all} \; x \geq 0 \right\}
\end{align*}
where for every $x \geq 0$
%\begin{align*}
%F(x) = \frac {1-A(x)} {W^{(q)'}(x)}
%\end{align*}
\begin{align}
F(x) := \frac {1-A(x)} {W^{(q)'}(x)},  \label{F}
\end{align}
and
\begin{align*}
A(x) := K \left( Z^{(q)}(x) - \psi'(0+) W^{(q)}(x) \right) + S q W^{(q)} (x).
\end{align*}
The value function is given by $\widetilde{v}_{b^*}(x)$ with
\begin{align}
\widetilde{v}_{b}(x) = \left\{ \begin{array}{ll}S + \int_0^x A(y) \diff y + \frac {1-A(b)} {W^{(q)'}(b)} W^{(q)}(x), & 0 \leq x \leq b, \\ x - b +v_{b}(b) , & x > b.\end{array} \right. \label{v_tilde}
\end{align}

\textbf{Extension with transaction costs.}
An extension allowing transaction costs is discussed in \cite{Loeffen_2009_2} where the objective is to maximize
\begin{align*}
\E^x \left[ \int_0^{\sigma^\pi} e^{-qt} \diff \left( L_t^\pi - \sum_{0 \leq s < t} \delta 1_{\{ \Delta L_s^\pi > 0 \}}\right)\right].
\end{align*}
Here $\delta > 0$ is the unit transaction cost and the strategy is assumed to be defined by a pure jump process in the form
\begin{align*}
L_t^\pi = \sum_{0 \leq s < t} \Delta L_s^\pi, \quad t \geq 0.
\end{align*}
For this \emph{impulse control} problem, the role of the barrier strategy in the classical model is now replaced by the so-called $(c_1,c_2)$-policy which is commonly known in inventory control.  The $(c_1,c_2)$-policy brings the risk process down to the level $c_1$ whenever the risk process goes above the level $c_2$. \cite{Loeffen_2009_2} showed that $(c^*_1,c^*_2)$-policy is optimal if it satisfies $c^*_2-c^*_1-\delta \geq 0$, minimizes the function
\begin{align*}
g(c_1,c_2) = \frac {W^{(q)}(c_2)-W^{(q)}(c_1)} {c_2-c_1-\delta},
\end{align*}
and satisfies $W^{(q)'}(a) \leq W^{(q)'}(b)$ for every $c_2^* \leq a \leq b$. The value function has the form (\ref{value_original}) where the optimal threshold level is replaced with $c_2^*$.

\section{Scale functions for spectrally negative \lev processes with phase-type jumps} \label{section_phase_type}
As we have seen in the last section, the implementation of the optimal strategies in the optimal dividend problem draws upon the computation of scale functions. This section obtains the scale function of the spectrally negative \lev process with phase-type jumps and shows that it can be used to approximate the scale function of any spectrally negative \lev process.

\subsection{Spectrally negative \lev processes with phase-type jumps}

Consider a continuous-time Markov chain $Y = \{ Y_t; t \geq 0 \}$ with finite
state space $\{1,\ldots,m \} \cup \{ \Delta \}$ where $1,\ldots,m$
are transient and $\Delta$ is absorbing. Its initial distribution is given by
a simplex ${\bm \alpha}=[\alpha_1, \ldots, \alpha_m]$
such that $\alpha_i=\p \left\{ Y_0=i \right\}$ for every $i = 1,\ldots,m$.  The intensity matrix ${\bm Q}$ is partitioned into the $m$ transient
states and the absorbing state $\Delta$, and is given by
\begin{align*}
{\bm Q}  := \begin{bmatrix} {\bm T} & {\bm t} \\ {\bm 0} & 0 \end{bmatrix}.
\end{align*}
Here  ${\bm T}$  is an  $m \times m$-matrix called the phase-type
generator, and ${\bm t} = - {\bm T} {\bm 1}$ where  ${\bm 1} =
[1,\ldots,1]'$. A distribution is called \emph{phase-type} with
representation $(m, {\bm \alpha}, {\bm T})$ if it is the
distribution of the absorption time to $\Delta$ in the Markov chain
described above. It is known that ${\bm T}$ is non-singular and thus
invertible; see \cite{Asmussen_1996}.  Its distribution and density functions
are given, respectively, by
\begin{align*}
F(z) =  1-{\bm \alpha} e^{{\bm T} z} {\bm 1} \quad \textrm{and} \quad f(z) = {\bm \alpha} e^{{\bm T} z} {\bm t}, \quad z \geq 0.
\end{align*}

Let $X = \left\{X_t; t \geq 0 \right\}$ be a \emph{spectrally negative} \lev process of the form
\begin{equation}
  X_t  - X_0=\mu t+\sigma B_t - \sum_{n=1}^{N_t} Z_n, \quad 0\le t <\infty, \label{levy_canonical}
\end{equation}
for some $\mu \in \R$ and $\sigma \geq 0$.  Here $B=\{B_t; t\ge 0\}$ is a standard Brownian motion, $N=\{N_t; t\ge 0\}$ is a Poisson process with arrival rate $\lambda$, and  $Z = \left\{ Z_n; n = 1,2,\ldots \right\}$ is an i.i.d.\ sequence of phase-type distributed random variables with representation $(m,{\bm \alpha},{\bm T})$. These processes are assumed independent. Its Laplace exponent is then
\begin{align}
 \psi(s)   = \mu s + \frac 1 2 \sigma^2 s^2 + \lambda \left( {\bm \alpha} (s {\bm I} - {\bm{T}})^{-1} {\bm t} -1 \right), \label{laplace_exponent_phase_type}
 \end{align}
which is analytic for every $s \in \mathbb{C}$ except for the eigenvalues of ${\bm T}$.

Disregarding the  case when $X$ is a negative subordinator, we consider the following two cases:
\begin{enumerate}
\item[] \textbf{Case 1:} when $\sigma > 0$ (i.e.\ $X$ has unbounded variation),
\item[]  \textbf{Case 2:} when $\sigma = 0$ and $\mu > 0$ (i.e.\ $X$ is a compound Poisson process).
\end{enumerate}
Notice, in Case 2, that we can write $X_t = U_t - \sum_{n=1}^{N_t}
Z_n$ where $U_t = x + \mu t$ is a (positive) subordinator.  This implies that
down-crossing of a threshold can occur only by jumps; see, for
example, Chapter III of \cite{Bertoin_1996}.  On the other
hand, in Case 1, down-crossing can occur also by \emph{creeping
downward} (by the diffusion components). Due to this difference, the form of the scale function differs as we shall see.

Fix $q > 0$. Consider the \emph{Cram\'{e}r-Lundberg} equation
\begin{align} \label{cramer_lundberg}
\psi(s) = q,
\end{align}
and define the set of (the absolute values of) \emph{negative roots} and the set of \emph{poles}:
\begin{align*}
\mathcal{I}_q &:= \left\{ i: \psi (-\xi_{i,q}) = q \; \textrm{and} \; \mathcal{R} (\xi_{i,q}) > 0\right\}, \\
\mathcal{J}_q &:= \left\{ j: \frac q {q - \psi(-\eta_j)} = 0 \; \textrm{and} \; \mathcal{R} (\eta_j) > 0\right\}.
\end{align*}
The elements in $\mathcal{I}_q$ and $\mathcal{J}_q$ may not be
distinct, and, in this case, we take each as many times as its
multiplicity. By Lemma 1 of
\cite{Asmussen_2004}, we have
\begin{align*}
|\mathcal{I}_q| = \left\{ \begin{array}{ll} |\mathcal{J}_q| + 1, & \textrm{for Case 1}, \\ |\mathcal{J}_q|, & \textrm{for Case 2}. \end{array} \right.
\end{align*}
In particular, if the representation is minimal (see \cite{Asmussen_2004}), we have $|\mathcal{J}_q| = m$.

Let $\kappa_q$ be an independent exponential random variable with parameter $q$ and denote the \emph{running maximum} and \emph{minimum}, respectively, by
\begin{align*}
\overline{X}_t = \sup_{0 \leq s \leq t} X_s \quad \textrm{and} \quad \underline{X}_t = \inf_{0 \leq s \leq t} X_s, \quad t \geq 0.
\end{align*}
The \emph{Wiener-Hopf factorization} states that $q /{(q - \psi(s))} = \varphi_q^+ (s) \varphi_q^- (s)$ for every $s \in \mathbb{C}$ such that $\mathcal{R}(s) = 0$, with the
\emph{Wiener-Hopf factors}
\begin{align}
\varphi_q^- (s) := \E \left[ \exp (s \underline{X}_{\kappa_q}) \right] \quad \textrm{and} \quad \varphi_q^+ (s) := \E \left[ \exp (s \overline{X}_{\kappa_q}) \right] \label{wiener_hopf}
\end{align}
that are analytic for $s$ with $\mathcal{R}(s) > 0$ and $\mathcal{R}(s) < 0$, respectively. By Lemma 1 of \cite{Asmussen_2004}, we have, for every $s$ such that $\mathcal{R}(s) > 0$,
\begin{align}
\varphi_q^- (s) = \frac {\prod_{j \in \mathcal{J}_q} (s+\eta_j)} {\prod_{j \in \mathcal{J}_q} \eta_j} \frac {\prod_{i \in \mathcal{I}_q} \xi_{i,q}} {\prod_{i \in \mathcal{I}_q} (s+\xi_{i,q})}, \label{winer_hopf_phase_type}
\end{align}
from which we can obtain the distribution of
$\underline{X}_{\kappa_q}$ by the Laplace inverse via partial
fraction expansion.

As in Remark 4 of \cite{Asmussen_2004}, let
$n$ denote the number of different roots in $\mathcal{I}_q$ and
$m_i$ denote the multiplicity of a root $\xi_{i,q}$ for $i = 1,\ldots,n$. Then we have
\begin{align}
\mathbb{P} \left\{-\underline{X}_{\kappa_q} \in \diff x \right\} = \sum_{i = 1}^n \sum_{k=1}^{m_i} A_{i,q}^{(k)} \xi_{i,q} \frac {(\xi_{i,q} x)^{k-1}} {(k-1)!} e^{-\xi_{i,q} x} \diff x, \quad x > 0 \label{dist_x_kappa_2}
\end{align}
where
\begin{align*}
A_{i,q}^{(k)} := \left. \frac 1 {(m_i-k)!} \frac {\partial^{m_i-k}} {\partial s^{m_i-k}} \frac {\varphi^-_q(s) (s+\xi_{i,q})^{m_i}} {\xi_{i,q}^k} \right|_{s = -\xi_{i,q}}.
\end{align*}
Notice that this can be simplified significantly when all the roots in $\mathcal{I}_q$ are distinct.

\subsection{Scale functions for spectrally negative \lev processes with phase-type jumps}
Here we obtain the scale function.  We focus on the case $q > 0$ because the scale
function when $q=0$ (and $X$ drifts to infinity) can be derived by using $W^{(0)}(x) = \mathbb{P}^x \left\{ \underline{X}_\infty \geq 0 \right\}/\psi'(0)$ and the ruin probability (19) of \cite{Asmussen_2004} by taking $q \rightarrow 0$. \cite{Kyprianou_Palmowski_2007} briefly stated the scale
function when $q=0$ and all the roots in $\mathcal{I}_q$ are distinct.

Before obtaining the scale function, we shall first represent the positive
root $\zeta_q$ (\ref{zeta}) in terms of the negative roots $\left\{ \xi_{i,q};\; i \in
\mathcal{I}_q \right\}$. Let us define
\begin{align}
\varrho_q :=  \sum_{i = 1}^n A_{i,q}^{(1)} \xi_{i,q}, \quad q > 0, \label{eq:varrho1}
\end{align}
and by Lemma \ref{lemma_zero}
\begin{align}\label{eq:theta}
\theta := -\zeta_q W^{(q)} (0) + W^{(q)'} (0+) = \left\{ \begin{array}{ll}  \frac 2 {\sigma^2}, & \textrm{for Case 1} \\ - \frac {\zeta_q} {\mu} + \frac {q + \lambda} {\mu^2}, & \textrm{for Case 2} \end{array} \right\}.
\end{align}
%We defer the proofs of Lemmas \ref{lemma_zeta_wrt_xi} and \ref{lemma_scale_function} to the appendix.
\begin{lemma} \label{lemma_zeta_wrt_xi}
For every $q > 0$, we have
\begin{align*}
\frac {\zeta_q} q = \frac \theta {\varrho_q}.
\end{align*}

\end{lemma}

We now obtain the version of the scale function $W_{\zeta_q} (\cdot)$.  In the lemma below, $W_{\zeta_q}(0) = W^{(q)} (0)$ is either $0$ or $\frac 1 \mu$ depending on if it is Case 1 or Case 2; see Lemma \ref{lemma_zero}.

\begin{lemma} \label{lemma_scale_function}
For every $q > 0$, we have
\begin{align*}
W_{\zeta_q}(x) - W_{\zeta_q}(0) =  \frac {\zeta_q} q \sum_{i = 1}^n \sum_{k=1}^{m_i} A_{i,q}^{(k)} \left( \frac {\xi_{i,q}} {\zeta_q + \xi_{i,q}} \right)^{k} \left[ 1 - e^{-(\zeta_q + \xi_{i,q})x} \sum_{j=0}^{k-1} \frac {((\zeta_q + \xi_{i,q})x)^j} {j!} \right], \quad x \geq 0.
\end{align*}

\end{lemma}

Lemma \ref{lemma_scale_function}
together with (\ref{scale_function_asymptotic}) and Lemmas \ref{lemma_zero} and \ref{lemma_zeta_wrt_xi} shows the following.

\begin{proposition} \label{proposition_scale_function}
For every $q > 0$ and $x \geq 0$, we have the following.
\begin{enumerate}
\item For Case 1, we have
\begin{align*}
W^{(q)}(x) =  \frac 2 {\sigma^2 \varrho_q} \sum_{i = 1}^n \sum_{k=1}^{m_i} A_{i,q}^{(k)} \left( \frac {\xi_{i,q}} {\zeta_q + \xi_{i,q}} \right)^{k} \left[ e^{\zeta_q x} - e^{-\xi_{i,q}x} \sum_{j=0}^{k-1} \frac {((\zeta_q + \xi_{i,q})x)^j} {j!} \right].
\end{align*}

\item For Case 2, we have
\begin{align*}
W^{(q)}(x) =  \frac 1 {\varrho_q} \left( - \frac {\zeta_q} {\mu} + \frac {q + \lambda} {\mu^2} \right) \sum_{i = 1}^n \sum_{k=1}^{m_i} A_{i,q}^{(k)} \left( \frac {\xi_{i,q}} {\zeta_q + \xi_{i,q}} \right)^{k} \left[ e^{\zeta_q x} - e^{- \xi_{i,q}x} \sum_{j=0}^{k-1} \frac {((\zeta_q + \xi_{i,q})x)^j} {j!} \right]  +\frac 1 {\mu} e^{\zeta_q x}.
\end{align*}

\end{enumerate}
\end{proposition}
Recall that the solution to the dividend problem requires the derivative. The scale functions obtained above are infinitely differentiable.  In particular, the first derivative becomes
\begin{align*}
W^{(q)'}(x)
=  \frac 2 {\sigma^2 \varrho_q} \sum_{i = 1}^n \sum_{k=1}^{m_i} A_{i,q}^{(k)} \left( \frac {\xi_{i,q}} {\zeta_q + \xi_{i,q}} \right)^{k} \left[ \zeta_q e^{\zeta_q x} + \xi_{i,q} e^{-\xi_{i,q}x} \frac {((\zeta_q + \xi_{i,q})x)^{k-1}} {(k-1)!} - \zeta_q e^{-\xi_{i,q}x} \sum_{j=0}^{k-2} \frac {((\zeta_q + \xi_{i,q})x)^j} {j!} \right]
\end{align*}
for Case 1 and
\begin{multline*}
W^{(q)'}(x) =  \frac 1 {\varrho_q} \left( - \frac {\zeta_q} {\mu} + \frac {q + \lambda} {\mu^2} \right) \sum_{i = 1}^n \sum_{k=1}^{m_i} A_{i,q}^{(k)} \left( \frac {\xi_{i,q}} {\zeta_q + \xi_{i,q}} \right)^{k} \\ \times \left[ \zeta_q e^{\zeta_q x} + \xi_{i,q} e^{-\xi_{i,q}x} \frac {((\zeta_q + \xi_{i,q})x)^{k-1}} {(k-1)!} - \zeta_q e^{-\xi_{i,q}x} \sum_{j=0}^{k-2} \frac {((\zeta_q + \xi_{i,q})x)^j} {j!} \right]  +\frac 1 {\mu} \zeta_q e^{\zeta_q x}
\end{multline*}
for Case 2.

When all the roots in $\mathcal{I}_q$ are distinct, the scale functions above can be simplified and have nice properties as discussed in the following corollary.
\begin{corollary} \label{remark_distinct}
If all the roots in $\mathcal{I}_q$ are distinct, we have the followings. 
\begin{enumerate}
\item The scale function can be simplified to
\begin{align*}
W^{(q)}(x) &=  \frac 2 {\sigma^2 \varrho_q} \sum_{i = 1}^n A_{i,q}^{(1)} \left( \frac {\xi_{i,q}} {\zeta_q + \xi_{i,q}} \right) \left[ e^{\zeta_q x} - e^{-\xi_{i,q}x} \right], \\
W^{(q)}(x) &=  \frac 1 {\varrho_q} \left( - \frac {\zeta_q} {\mu} + \frac {q + \lambda} {\mu^2} \right) \sum_{i = 1}^n A_{i,q}^{(1)} \left( \frac {\xi_{i,q}} {\zeta_q + \xi_{i,q}} \right) \left[ e^{\zeta_q x} - e^{- \xi_{i,q}x} \right]  +\frac 1 {\mu} e^{\zeta_q x},
\end{align*}
for Case 1 and Case 2, respectively.
\item $W^{(q)'}$ is convex.
\item $W_{\zeta_q}'$ is completely monotone.
\end{enumerate}
\end{corollary}
This guarantees the optimality of the barrier-strategy in view of Theorem \ref{optimality_barrier}; the optimal strategy can be obtained by finding a unique $a^*$ such that $W^{(q)''}(a^*)=0$.

\begin{example}[Hyperexponential Case] \label{example_hyperexponential}
As an important example where all the roots in $\mathcal{I}_q$ are distinct, we consider the case where $Z$ has a hyperexponential distribution with density function
\begin{align*}
f (z)  = \sum_{i=1}^m \alpha_i \eta_i e^{- \eta_i z}, \quad z \geq 0,
\end{align*}
for some $0 < \eta_1 < \cdots < \eta_m < \infty$.  Its Laplace exponent (\ref{laplace_spectrally_negative})
is then
\begin{align}\label{laplace_exponent_double_exponential}
\psi(s) = \mu s + \frac 1 2 \sigma^2 s^2  - \lambda
\sum_{i=1}^m \alpha_i \frac s {\eta_i + s}.
\end{align}
Notice in this case that $-\eta_1$, \ldots, $-\eta_m$ are the poles of the Laplace exponent.  Furthermore, all the roots in $\mathcal{I}_q$ are distinct and satisfy the following interlacing condition for every $q > 0$:
\begin{enumerate}
\item
when $\sigma > 0$, there are $m+1$ roots $-\xi_{1,q}, \ldots, -\xi_{m+1,q}$ such that
\begin{align*}
0 < \xi_{1,q} < \eta_1 < \xi_{2,q} < \cdots < \eta_m < \xi_{m+1,q} < \infty;
\end{align*}
\item when $\sigma = 0$ and $\mu > 0$, there are $m$ roots $-\xi_{1,q}, \ldots, -\xi_{m,q}$  such that
\begin{align*}
0 < \xi_{1,q} < \eta_1 < \xi_{2,q} < \cdots <  \xi_{m,q}  < \eta_m < \infty.
\end{align*}
\end{enumerate}
The class of hyperexponential distributions is important as it is dense in the class of all positive-valued distributions with completely monotone densities.
\end{example}

\subsection{Approximation of the scale function of a general spectrally negative \lev process} \label{section_approximation}
The scale function obtained in Proposition
\ref{proposition_scale_function} can be used to approximate the
scale function of a general spectrally negative \lev process.  By
Proposition 1 of \cite{Asmussen_2004}, there exists,
for any spectrally negative \lev process $X$, a sequence of
spectrally negative \lev processes with phase-type jumps  $X^{(n)}$
converging to $X$ in $D[0,\infty)$.  This is equivalent to saying
that $X_1^{(n)} \rightarrow X_1$ in distribution by \cite{Jacod_Shirayev_2003}, Corollary VII 3.6; see also \cite{Pistorius_2006}. Suppose $\psi_n$ ($\psi$), $\zeta_{q,n}$ ($\zeta_q$) and
$W^{(q)}_n/W_{\zeta_q,n}$ ($W^{(q)}/W_{\zeta_q}$) are the Laplace exponent, the positive root (\ref{zeta}) and the scale
function of $X^{(n)}$ ($X$), respectively.  Because these processes
are spectrally negative and $\psi$ is continuous, we have,  by the continuity theorem,
$\psi_n(\beta + \zeta_{q,n}) \rightarrow \psi(\beta + \zeta_q)$ for every $\beta > 0$. Now in view of
(\ref{scale_version}), the convergence of the scale function holds by the
continuity theorem; see \cite{Feller_1971}, Theorem 2a,
XIII.1.  More precisely, we have $\int_I W_{\zeta_q,n} (y) \diff y
\rightarrow \int_I W_{\zeta_q} (y) \diff y$ and $\int_I W^{(q)}_n (y) \diff y
\rightarrow \int_I W^{(q)} (y) \diff y$ for any interval $I$.

The smoothness and monotonicity properties of the scale function can be additionally used to obtain stronger results. The scale functions in Proposition \ref{proposition_scale_function} are in $C^\infty(0,\infty)$. In addition, when all the roots of $\mathcal{I}_q$ are different, its first derivative $W_{\zeta_q}'$ is completely monotone as discussed in Corollary \ref{remark_distinct}.
%By Chan et al.\ \cite{Chan_2009}, it is in  $ C^2(0,\infty)$ whenever there is a Gaussian component.   %As shown by Loeffen \cite{Loeffen_2008}, if the jump distribution has a completely monotone density, it is in $C^\infty(0,\infty)$.

If the target scale function is in $C^1(0,\infty)$ (which holds whenever the jump distribution has no atoms), noting that
$W_{\zeta_q}(x) \leq (\psi'(\zeta_q))^{-1}$ for every $x$ and hence
$e^{-\beta x}W_{\zeta_q}(x)$ vanishes in the limit for any $\beta >
0$, we have by (\ref{scale_version})
\begin{align*}
\int_0^\infty e^{-\beta x} W'_{\zeta_q} (x) \diff x &= \frac \beta {\psi(\beta+\zeta_q)-q} - W_{\zeta_q} (0), \quad \beta > 0.
\end{align*}
Because $W'_{\zeta_q} (x)$ is nonnegative and $F(x)  :=  \int_0^x
W'_{\zeta_q} (y) \diff y / ((\psi'(\zeta_q))^{-1} - W_{\zeta_q}
(0))$ is a probability distribution,
\begin{align*}
 W_{\zeta_q,n} (x) \xrightarrow{n \uparrow \infty}   W_{\zeta_q} (x)  \quad  \textrm{and} \quad W^{(q)}_n (x)  \xrightarrow{n \uparrow \infty}  W^{(q)} (x), \quad x \geq 0.
\end{align*}

Furthermore, suppose that it is
in $C^2(0,\infty)$ (which holds, for example, when $\sigma > 0$  by \cite{Chan_2009}), $W'_{\zeta_q} (0+) < \infty$ (i.e., $\sigma > 0$
or $\Pi(-\infty,0) < \infty$) and $W''_{\zeta_q} (x) \leq 0$ for
every $x \geq 0$, because $W'_{\zeta_q}(x) \xrightarrow{x \uparrow
\infty} 0$, we have $F(x) := (W'_{\zeta_q} (0+))^{-1} \int_0^x
|W''_{\zeta_q} (y)| \diff y$ is a probability distribution and
\begin{align*}
\int_0^\infty e^{-\beta x} F (\diff x) &= (W'_{\zeta_q} (0+))^{-1} \left[- \frac {\beta^2} {\psi(\beta+\zeta_q)-q} + \beta W_{\zeta_q} (0) + W'_{\zeta_q} (0+) \right], \quad \beta > 0.
\end{align*}
Therefore, noting that $W^{(q)'} (x) = \zeta_q W^{(q)} (x) +
e^{\zeta_q x} W'_{\zeta_q} (x)$ and assuming that the convergent
sequence $W_{\zeta_q,n} (x)$ has the same property, we can obtain by
the continuity theorem
\begin{align*}
 W'_{\zeta_q,n} (x)  \xrightarrow{n \uparrow \infty}   W'_{\zeta_q} (x)  \quad  \textrm{and} \quad W^{(q)'}_n (x)  \xrightarrow{n \uparrow \infty}  W^{(q)'} (x), \quad x \geq 0.
\end{align*}
The negativity of $W''_{\zeta_q}$ holds, for example, for the completely monotone jump case because $W'_{\zeta_q}$ is completely monotone by \cite{Loeffen_2008}. We can also choose the sequence $W_{\zeta_q,n}'$ completely monotone in view of Corollary \ref{remark_distinct} because approximation can be done via hyperexponential distributions.  In fact, it also means that $W_{\zeta_q}$ is $C^{\infty}(0,\infty)$ and the convergence of higher derivatives can be pursued. Even for a general jump distribution, the negativity  of $W''_{\zeta_q}$ is a reasonable assumption in view of the numerical plots given by \cite{Surya_2008}.

This phase-type fitting approach complements the approach by \cite{Surya_2008} where scale functions are approximated by numerical Laplace inversion.  A major disadvantage of using this inversion method is the fact that it requires the exact value of the right-hand side of (\ref{eq:scale}).  However, the Laplace transform of a jump distribution does not in general have an explicit closed-form expression. \cite{Surya_2008}'s approach, therefore, contains two types of errors: 1) the approximation error caused while computing $\psi$ and 2) the error caused while inverting the Laplace transform.  On the other hand, the phase-type fitting approach only contains the phase-type fitting error thanks to the closed-form Laplace transform of the phase-type distribution.

The phase-type fitting approach enjoys a variety of fitting algorithms typically developed in queueing analysis.
Well-known examples are the moment-matching approach (e.g.\ MEFIT and MEDA) and the maximum-likelihood approach (e.g.\ MLAPH and EMPHT), and a thorough study of pros and cons of each fitting techniques has been conducted in, for example, \cite{Horvath_Telek} and \cite{Lang_Arthur_1996}.  The fitting can be applied also to empirical data and this is another major advantage over the Laplace inversion approach.

\section{Scale functions for Meromorphic \lev processes} \label{section_meromorphic}

In this section, we consider another class of spectrally negative \lev processes called meromorphic \lev processes.  We obtain their scale functions and use these as approximation tools for a general spectrally negative \lev process with a completely monotone \lev measure.  Similarly to the approach applied in the last section, we obtain the scale function using its Wiener-Hopf factorization.  It has a form expressed as an infinite sum of exponential functions which can be bounded efficiently by finite sums.
\subsection{Meromorphic \lev processes}
The following is due to \cite{Kuznetsov_2010}, Definition 1.

\begin{definition}[spectrally negative meromorphic \lev process] \label{definition_meromorphic}
A spectrally negative \lev process $X$ is said to belong to the M-class if the following conditions hold.
\begin{enumerate}
\item The Laplace exponent $\psi(s)$ (\ref{laplace_spectrally_negative}) has a countable set of real negative poles.
\item For every $q \geq 0$, the Cram\'er-Lundberg equation (\ref{cramer_lundberg}) has a countable set of real negative roots.
\item Let $\left\{ \eta_k; k \geq 1 \right\}$ and $\left\{ \xi_{k,q}; k \geq 1 \right\}$, respectively, be the sets of the absolute values of the poles and the negative roots of (\ref{cramer_lundberg}) for fixed $q \geq 0$. Then it satisfies the following interlacing conditions:
\begin{align*}
\cdots < -\eta_k < -\xi_{k,q} < \cdots < -\eta_2 < -\xi_{2,q} < -\eta_1 < -\xi_{1,q} < 0.
\end{align*}
\item There exists $\alpha > \frac 1 2$ such that $\eta_k \sim c k^\alpha$ as $k \rightarrow \infty$.
\item The Wiener-Hopf factor (\ref{wiener_hopf}) is expressed as convergent infinite products
\begin{align}
\varphi_q^- (s) = \prod_{k = 1}^\infty \frac { (s+\eta_k)} {\eta_k} \frac { \xi_{k,q}} {(s+\xi_{k,q})}. \label{wiener_hopf_meromorphic}
\end{align}
\end{enumerate}
\end{definition}

The M-class complements the class of \lev processes with phase-type
jumps described in the previous section because it also contains those of infinite activity.  As noted by Corollary 3 of \cite{Kuznetsov_2010},
the property (3) in Definition 4.1 is equivalent to the condition that the \lev measure has the form
(\ref{density_meromorphic}). This can be seen as an
extension to the hyperexponential case as described in Example
\ref{example_hyperexponential}. 

We consider the M-class
alternatively to the ``hyperexponential fitting".  When a \lev measure is
completely monotone, approximation via hyperexponential distributions is in
principle possible.  However, as in, for example, \cite{AsmussenMadanPistorius07}, special care is needed for the infinitesimal jumps, and one needs to approximate separately the process with \lev measure $\nu(-\varepsilon,0)$ for small $\varepsilon > 0$.  Fitting via the M-class is more tractable in the sense that this procedure is not necessary.  Although
property (4) requires one to choose $\eta$'s in a certain way, the
approximation for the \lev process with a completely monotone
density is still effective by choosing the value of $c$ and $\alpha$
sufficiently small.   For more details, see \cite{Kuznetsov_2010}.

The Wiener-Hopf factor (\ref{wiener_hopf_meromorphic}) is again a rational function as in (\ref{winer_hopf_phase_type}) for the phase-type case.  Therefore, this can be inverted again by partial fraction decomposition, and we have

\begin{align}
\mathbb{P} \left\{ -\underline{X}_{\kappa_q} \in \diff x \right\} =  \sum_{k=1}^\infty A_{k,q} \xi_{k,q} e^{-\xi_{k,q} x}  \diff x, \quad x > 0 \label{dist_x_kappa}
\end{align}
where
\begin{align*}
A_{k,q} := \left. \frac {s+\xi_{k,q}} {\xi_{k,q}} \varphi_q^- (s) \right|_{s=-\xi_{k,q}} = \left( 1 - \frac {\xi_{k,q}} {\eta_k} \right) \prod_{i \neq k} \frac {1 - \frac {\xi_{k,q}} {\eta_i}} {1 - \frac {\xi_{k,q}} {\xi_{i,q}}}, \quad k \geq 1.
\end{align*}
Notice by the interlacing condition that $A_{k,q} > 0$ for every $k \geq 1$.

\subsection{Scale functions for meromorphic \lev processes}
We now obtain the scale function for the M-class.  We omit the proof
because it is similar to the phase-type case; see Appendix
\ref{subsection_proof}.

\begin{lemma} \label{lemma_scale_function_meromorphic}
For every $q > 0$, we have
\begin{align*}
W_{\zeta_q}(x) - W_{\zeta_q}(0) =  \sum_{k=1}^\infty C_{k,q} \left[ 1- e^{-(\zeta_q+\xi_{k,q}) x} \right], \quad x \geq 0
\end{align*}
where
\begin{align}
C_{k,q} := \frac {\zeta_q} q \frac {\xi_{k,q}A_{k,q}} {\zeta_q+\xi_{k,q}}, \quad k \geq 1. \label{C}
\end{align}
\end{lemma}
By (\ref{scale_function_asymptotic2}) and Lemma \ref{lemma_scale_function_meromorphic}, we have, by taking the limit,
\begin{align}
\kappa_q := \sum_{k=1}^\infty C_{k,q} = (\psi'(\zeta_q))^{-1} - W_{\zeta_q}(0)  < \infty.  \label{eq_varrho}
\end{align}
The scale function can be therefore obtained by Lemma \ref{lemma_scale_function_meromorphic} and \eqref{eq_varrho}.
\begin{proposition} \label{proposition_scale_function2}For every $q > 0$, we have
\begin{align} \label{scale_function_before}
W^{(q)}(x)  = \sum_{i=1}^\infty C_{i,q} \left[ e^{\zeta_q x}- e^{-\xi_{i,q} x} \right] +  W_{\zeta_q}(0)  e^{\zeta_q x} = (\psi'(\zeta_q))^{-1} e^{\zeta_q x} - \sum_{i=1}^\infty C_{i,q}e^{-\xi_{i,q} x}, \quad x \geq 0.
\end{align}
\end{proposition}

By straightforward differentiation, we have, for every $q > 0$ and $x \geq 0$,
\begin{align*}
W^{(q)'}(x)  &= (\psi'(\zeta_q))^{-1} \zeta_q e^{\zeta_q x} + \sum_{i=1}^\infty C_{i,q}\xi_{i,q}e^{-\xi_{i,q} x}, \\
W^{(q)''}(x)  &= (\psi'(\zeta_q))^{-1} (\zeta_q)^2 e^{\zeta_q x} - \sum_{i=1}^\infty C_{i,q} (\xi_{i,q})^2 e^{-\xi_{i,q} x}.
%W^{(q)'''}(x)
%&= (\psi'(\zeta_q))^{-1} (\zeta_q)^3 e^{\zeta_q x} + \sum_{i=1}^\infty C_{i,q}(\xi_{i,q})^3 e^{-\xi_{i,q} x}.
\end{align*}
\begin{remark} \label{remark_optimal_level}
From the derivatives above, it can be easily verified that
$W^{(q)'}(\cdot)$ is indeed convex. In view of Theorem
\ref{theorem_optimality}, the optimal solution to the classical
dividend problem (\ref{eq:classical-p}) is the unique point $x$
such that $W^{(q)''}(x)$ vanishes or
\begin{align*}
(\psi'(\zeta_q))^{-1} (\zeta_q)^2 e^{\zeta_q x} - \sum_{i=1}^\infty C_{i,q} (\xi_{i,q})^2 e^{-\xi_{i,q} x} = 0.
\end{align*}
\end{remark}
As an extension to Lemma \ref{lemma_zeta_wrt_xi}, we have the following.
\begin{lemma} \label{lemma_zeta_wrt_xi_meromorphic}
\begin{enumerate}
\item The following two statements are equivalent:
\begin{enumerate}
\item $\sigma = 0$ and $\nu(-\infty,0) = \infty$,
\item $\sum_{k=1}^\infty A_{k,q} \xi_{k,q} = \infty$.
\end{enumerate}
\item Suppose $\sigma > 0$ or $\nu(-\infty,0) < \infty$.  Then, for every $q > 0$, we have
\begin{align} \label{zeta_divided_by_q}
\frac {\zeta_q} q = \theta \left(\sum_{k=1}^\infty A_{k,q} \xi_{k,q}\right)^{-1}
\end{align}
where
\begin{align}
\theta := -\zeta_q W^{(q)} (0) + W^{(q)'} (0+) = \left\{ \begin{array}{ll}  \frac 2 {\sigma^2}, & \textrm{when $\sigma > 0$} \\ - \frac {\zeta_q} {\mu} + \frac {q + \nu(-\infty,0)} {\mu^2}, & \textrm{when $\sigma =0$} \end{array} \right\}.
\end{align}
\end{enumerate}
\end{lemma}

\subsection{Approximation of the scale functions via finite sum}
 \label{section_approximation}
The scale function obtained in Proposition \ref{proposition_scale_function2} is an infinite sum of exponential functions and in reality its exact value cannot be computed. Here, we obtain bounds for $W^{(q)}(\cdot)$, $W^{(q)'}(\cdot)$ and $Z^{(q)}(\cdot)$ in terms of finite sums.

For every $m \geq 1$, let
\begin{align*}
A_{k,q}^{(m)} := 1_{\{ k \leq m\}}\left( 1 - \frac {\xi_{k,q}} {\eta_k}\right) \prod_{1 \leq i \leq m, i \neq k} \frac {1 - \frac {\xi_{k,q}} {\eta_{i}}} {1 - \frac {\xi_{k,q}}{\xi_{i,q}}} \quad \textrm{and} \quad C^{(m)}_{k,q} := \frac {\zeta_q} q \frac {\xi_{k,q}A^{(m)}_{k,q}} {\zeta_q+\xi_{k,q}}, \quad k \geq 1.
\end{align*}
By the interlacing condition, $A_{k,q}^{(m)}$ and $C^{(m)}_{k,q}$ are all positive and, for every $k \geq 1$,
\begin{align*}
A_{k,q}^{(m)} \uparrow A_{k,q} \quad \textrm{and} \quad C^{(m)}_{k,q} \uparrow C_{k,q} \textrm{ as } m \rightarrow \infty.
\end{align*}

Now we define candidates for upper and lower bounds of $W_{\zeta_q}$ respectively by
\begin{align*}
\overline{W}_{\zeta_q}^{(m)}(x) &:= (\psi'(\zeta_q))^{-1} - \sum_{i=1}^m C^{(m)}_{i,q}  e^{-(\zeta_q+\xi_{i,q}) x}, \\
\underline{W}_{\zeta_q}^{(m)}(x) &:= \overline{W}_{\zeta_q}^{(m)}(x) -\delta_m \left[ e^{-\zeta_q x} +  e^{-(\zeta_q+\xi_{m+1,q}) x} \right],
\end{align*}
for every $m \geq 1$ and $x \geq 0$, where
\begin{align*}
\delta_m := \kappa_q - \sum_{i=1}^m C_{i,q}^{(m)} > 0,
\end{align*}
which vanishes in the limit as $m \rightarrow \infty$ by \eqref{eq_varrho}.   As candidates for upper and lower bounds of $W^{(q)}$, we also define
\begin{align*}
\overline{W}^{(q,m)}(x) := e^{\zeta_q x} \overline{W}_{\zeta_q}^{(m)}(x)  \quad \textrm{and} \quad \underline{W}^{(q,m)}(x) := e^{\zeta_q x}\underline{W}_{\zeta_q}^{(m)}(x), \quad x \geq 0.
\end{align*}

The following proposition shows that the scale functions are bounded and approximated by these functions.
\begin{proposition} \label{proposition_bound_w}
For every $m \geq 1$ and $x \geq 0$, we have
\begin{align}
\underline{W}_{\zeta_q}^{(m)}(x) \leq   W_{\zeta_q}(x) \leq \overline{W}_{\zeta_q}^{(m)}(x) \quad \textrm{and} \quad \underline{W}^{(q,m)}(x) \leq W^{(q)}(x) \leq \overline{W}^{(q,m)}(x). \label{bounds_scale}
\end{align}
Furthermore, we have
\begin{align*}
\overline{W}_{\zeta_q}^{(m)}(x) \xrightarrow{m \uparrow \infty} W_{\zeta_q}(x) \quad \textrm{and} \quad \overline{W}^{(q,m)}(x) \xrightarrow{m \uparrow \infty} W^{(q)}(x), \\
\underline{W}_{\zeta_q}^{(m)}(x) \xrightarrow{m \uparrow \infty} W_{\zeta_q}(x) \quad \textrm{and} \quad \underline{W}^{(q,m)}(x) \xrightarrow{m \uparrow \infty} W^{(q)}(x),
\end{align*}
uniformly on $x \in [0,\infty)$.
\end{proposition}

By straightforward calculation, we can bound $Z^{(q)}$ in (\ref{def_z}). Let, for every $m \geq 1$, \begin{align*}
\overline{Z}^{(q,m)}(x) := 1 + q \int_0^x \overline{W}^{(m,q)}(y) \diff y \quad \textrm{and} \quad \underline{Z}^{(q,m)}(x) := 1 + q \int_0^x \underline{W}^{(m,q)}(y) \diff y, \quad x \geq 0.
\end{align*}
Then by Proposition \ref{proposition_bound_w}, we have $\underline{Z}^{(q,m)}(x) \leq Z^{(q,m)}(x) \leq \overline{Z}^{(q,m)}(x)$ and
\begin{multline*}
0 \leq \overline{Z}^{(q,m)}(x) - \underline{Z}^{(q,m)}(x) = q \int_0^x \left(\overline{W}^{(q,m)}(y)-\underline{W}^{(q,m)}(y) \right) \diff y \\ = q \delta_m \int_0^x \left[ 1+ e^{-\xi_{m+1,q} y} \right] \diff y = q \delta_m \left[ x + \frac 1 {\xi_{m+1,q}} (1- e^{-\xi_{m+1,q} x}) \right].
\end{multline*}
We therefore have the following.
\begin{corollary}[Bounds on $Z^{(q)}$]
We have $\overline{Z}^{(q,m)}(x) \rightarrow Z^{(q)}(x)$ and $\underline{Z}^{(q,m)}(x) \rightarrow Z^{(q)}(x)$ as $m \rightarrow \infty$ pointwise for every $x \geq 0$.
\end{corollary}

We now obtain bounds for the derivative. Define, for every $x > 0$,
\begin{align*}
\underline{w}^{(m)}(x) &:= (\psi'(\zeta_q))^{-1} \zeta_q e^{\zeta_q x} + \sum_{i=1}^m C_{i,q}^{(m)}\xi_{i,q}e^{-\xi_{i,q} x}, \\
\overline{w}^{(m)}(x) &:= \underline{w}^{(m)}(x) + \left[ \max_{1 \leq k \leq m} (\xi_{k,q}e^{-\xi_{k,q} x})  + \max_{k \geq m+1} (\xi_{k,q}e^{-\xi_{k,q} x})  \right] \delta_m.
\end{align*}
Here notice that
\begin{align*}
\max_{k \geq m+1} (\xi_{k,q}e^{-\xi_{k,q} x}) = 
\left\{ \begin{array}{ll} \frac 1 x, & \xi_{m+1,q} \leq \frac 1 x, \\ \xi_{m+1,q}e^{-\xi_{m+1,q} x}, & \xi_{m+1,q} > \frac 1 x. \end{array} \right.
\end{align*}

%\red{Should the last $\epsilon_m$ read $\delta_m$?}
\begin{proposition} \label{proposition_bounds_derivative}
For every $m \geq 1$, we have
\begin{align*}
\underline{w}^{(m)}(x)  \leq W^{(q)'}(x) \leq \overline{w}^{(m)}(x), \quad x \geq 0.
\end{align*}
Furthermore,  we have  $\underline{w}^{(m)}(x)  \rightarrow W^{(q)'}(x) $ and  $\overline{w}^{(m)}(x)  \rightarrow W^{(q)'}(x) $ uniformly on $x \geq x_0$ for any $x_0 > 0$.
\end{proposition}
A stronger result holds when $\sigma > 0$ or $\nu(-\infty,0) < \infty$.  Recall in this case that $\theta < \infty$ by Lemma  \ref{lemma_zeta_wrt_xi_meromorphic} (2) and hence we can define
\begin{align*}
\epsilon_m := \theta - \frac {\zeta_q} q \sum_{i=1}^m  \xi_{i,q} A_{i,q}^{(m)} > 0, \quad m \geq 1,
\end{align*}
which vanishes in the limit as $m \rightarrow \infty$ by Lemma \ref{lemma_zeta_wrt_xi_meromorphic} (2).

\begin{corollary} \label{corollary_z}
When  $\sigma > 0$ or $\nu(-\infty,0) < \infty$, we have
\begin{align*}
\underline{w}^{(m)}(x)  \leq W^{(q)'}(x) \leq  \overline{w}^{(m)}(x) \wedge \widetilde{w}^{(m)}(x), \quad x > 0
\end{align*}
where
\begin{align*}
\widetilde{w}^{(m)}(x) &:= \underline{w}^{(m)}(x) + \max_{1 \leq k \leq m} (\xi_{k,q}e^{-\xi_{k,q} x}) \delta_m + e^{-\xi_{m+1,q} x} \epsilon_m.
\end{align*}
\end{corollary}

The bounds obtained above on the derivative of the scale function can be used
to obtain bounds on the value functions and optimal barriers.  For
simplicity, let $\underline{w}$ and $\overline{w}$ be the lower and
upper bounds for $W^{(q)'}$ obtained above and let
\begin{align*}
w_* = \min_{x \geq 0}\underline{w} (x) \quad \textrm{and} \quad w^* = \min_{x \geq 0}\overline{w} (x).
\end{align*}
Clearly, $\underline{w}$ and $\overline{w}$ are convex because $\underline{w}^{(m)}$, $\overline{w}^{(m)}$ and $\widetilde{w}^{(m)}$ are for every fixed $m \geq 1$.

\begin{remark}[classical dividend problem]
Because  $\underline{w}$ and $\overline{w}$ are convex, we have $w_*
\leq W^{(q)'}(a^*)  \leq w^*$ and we have
\begin{align*}
\frac {\overline{W}^{(q,m)}(x) -\delta_m (1+e^{-\xi_{m,q} x})} {w^*} \leq u_{a^*}(x) \leq \frac {\overline{W}^{(q,m)}(x)} {w_*}.
\end{align*}
Furthermore, the optimal barrier must lie in
the following interval:
\begin{align*}
a^* \in \left\{ x \geq 0: \underline{w}(x) \leq w^*\right\}.
\end{align*}
\end{remark}

\begin{remark}[bail-out problem] For the bail-out problem,
recall that  the optimal barrier is the minimum value of $a$ such
that $G(a)$ becomes negative (see (\ref{optimal_barrier_bail_out})).
We can obtain its bounds simply by
\begin{align*}
\underline{G}(a) := [\varphi \underline{Z}^{(q,m)} (a) - 1 ] \underline{w}(a) - \varphi q \overline{W}^{(q,m)} (a)^2 \quad \textrm{and} \quad \overline{G}(a) &:= [\varphi \overline{Z}^{(q,m)} (a) - 1 ] \overline{w}(a) - \varphi q \underline{W}^{(q,m)} (a)^2.
\end{align*}
We can therefore obtain the bounds on the optimal barrier level $d^*$; we have $\underline{d}^* \leq d^* \leq \overline{d}^*$ where
\begin{align*}
\underline{d}^* := \inf \left\{a > 0: \underline{G}(a) \leq 0 \right\} \quad \textrm{and} \quad \overline{d}^* := \inf \left\{a > 0: \overline{G}(a) \leq 0 \right\}.
\end{align*}
\end{remark}

%\begin{remark}[with terminal value at ruin]
%By Lemma 4.2 of \cite{Loeffen_2010}, the function $F$ is increasing-decreasing on $(0,\infty)$ and this can be used to obtain bounds.
%We have $\underline{F}(x) \leq F(x) \leq \overline{F}(x)$ where
%\begin{align*}
%\overline{F}(x) &= (\underline{w}(x))^{-1}  \left[ 1 -  K \left( \underline{Z}^{(q,m)} (x) - \psi'(0+) \left[\overline{W}^{(q,m)} (x) 1_{\{\psi'(0+) \geq 0\}} + \underline{W}^{(q,m)} (x) 1_{\{\psi'(0+) < 0\}} \right] \right) - S q \underline{W}^{(q,m)} (x) \right], \\
%\underline{F}(x) &= (\overline{w}(x))^{-1}  \left[ 1 -  K \left( \overline{Z}^{(q,m)} (x) - \psi'(0+) \left[\underline{W}^{(q,m)} (x) 1_{\{\psi'(0+) \geq 0\}} + \overline{W}^{(q,m)} (x) 1_{\{\psi'(0+) < 0\}} \right] \right) - S q \overline{W}^{(q,m)} (x) \right].
%\end{align*}
%We then have
%\begin{align*}
%b^* \in \left\{ b \geq 0: \overline{F} (b) > \max_{y \geq 0} \underline{F} (b) \right\}.
%\end{align*}
%\end{remark}

\section{Numerical Examples} \label{section_numerical_results}
We conclude this paper by illustrating numerically the effectiveness of the phase-type fitting approximation for a general spectrally negative \lev process.  First, we use the classical hyperexponential fitting algorithm for a completely monotone density function by \cite{Feldmann_1998} and approximate, as an example, the scale function and solutions for the case with a (Brownian motion plus) compound Poisson with Weibull-distributed jumps.  Second, we consider, as an example of the meromorphic \lev process, the $\beta$-family introduced by \cite{Kuznetsov_2010_2} and extend the results to the spectrally negative version of the CGMY process.

\subsection{Brownian motion plus compound Poisson process with Weibull-distributed jumps}

As noted earlier, any spectrally negative \lev process with a completely monotone \lev measure can be approximated arbitrarily closely by fitting hyperexponential distributions.  Here, we use the fitted data computed by \cite{Feldmann_1998} to approximate the scale function when it is a Brownian motion plus a compound Poisson process with i.i.d.\ Weibull-distributed jumps.  Recall that the Weibull distribution with parameters $c$ and $a$ (Weibull($c$,$a$)) is give by
\begin{align*}
F(t) = 1-e^{-(t/a)^c}, \quad t \geq 0.
\end{align*}
%and the Pareto distribution with positive parameters $a$ and $b$ (denoted Pareto(a,b)) is given by
%\begin{align*}
%F(t) = 1-(1+bt)^{-a}, \quad t \geq 0.
%\end{align*}
If $c < 1$, it has long-tails, or $e^{\delta t} (1-F(t)) \rightarrow \infty$ as $t \rightarrow \infty$ for any $\delta > 0$, and has a completely monotone density.

%\cite{Feldmann_1998} showed that if a density function is \emph{completely monotone},
%then it can be approximated by those  of hyperexponential distributions. 
%As shown by \cite{Bernstein_1929}, every completely monotone density function is a mixture of exponential density functions, and this implies that, for any cdf with a completely monotone pdf, there exists a sequence of hyperexponential cdf's converging to it. The class of distributions with completely monotone densities contains a number of distributions such as the Pareto distribution, the Weibull distribution with $a < 1$, the gamma distribution with parameter less than $1$ and the Pareto mixture of exponentials distributions.  
%\cite{Feldmann_1998} took advantage of this fact and  proposed a recursive algorithm for fitting hyperexponential distributions to these distributions.

\cite{Feldmann_1998} constructed a recursive algorithm to approximate completely monotone densities in terms of hyperexponential densities.   We use their results and compute the scale functions of spectrally negative \lev processes with Weibull-distributed jumps. 

\begin{table}[ht]
\begin{tabular}{c}
\centering % used for centering table
\begin{tabular}{c c c | c c c} % centered columns (4 columns)
\hline\hline %inserts double horizontal lines
$i$ & $\alpha_i$ & $\eta_i$ & $i$ & $\alpha_i$ & $\eta_i$ \\ [0.5ex] % inserts table
%heading
\hline % inserts single horizontal line
1 & 0.029931 & 676.178 & 4 & 0.476233 & 0.76100\\ % inserting body of the table
2 & 0.093283 & 38.7090 & 5 & 0.068340 & 0.24800\\
3 & 0.332195 & 4.27400 & 6 & 0.000018 & 0.09700\\
%& &\\ [1ex] % [1ex] adds vertical space
\hline %inserts single line
\end{tabular} 
%& 
%\begin{tabular}{c c c} % centered columns (4 columns)
%\hline\hline %inserts double horizontal lines
%$i$ & $\alpha_i$ & $\eta_i$ \\ [0.5ex] % inserts table
%%heading
%\hline % inserts single horizontal line
%4 & 0.476233 & 0.76100 \\
%5 & 0.068340 & 0.24800 \\
%6 & 0.000018 & 0.09700 \\
%%& &\\ [1ex] % [1ex] adds vertical space
%\hline %inserts single line
%\end{tabular} 
%
%\begin{tabular}{c c c | c c c } % centered columns (4 columns)
%\hline\hline %inserts double horizontal lines
%$i$ & $\alpha_i$ & $\eta_i$ & $i$ & $\alpha_i$ & $\eta_i$\\ [0.5ex] % inserts table
%%heading
%\hline % inserts single horizontal line
%1 & 8.37E-11 & 8.3E-09 & 8 & 0.000147 & 0.0020  \\ % inserting body of the table
%2 & 7.18E-10 & 6.8E-08 & 9 & 0.001122 & 0.0100 \\
%3 & 5.56E-09 & 3.9E-07  & 10 & 0.008462 & 0.0570 \\
%4 & 4.27E-08 & 2.2E-06& 11 & 0.059768 & 0.3060 \\
%5 & 3.27E-07 & 1.2E-05  & 12 & 0.307218 & 1.5460 \\
%6 & 2.50E-06 & 6.5E-05  & 13 & 0.533823 & 6.5160 \\
%7 & 1.92E-05 & 3.5E-04  & 14 & 0.089437 & 23.304 \\ [1ex] % [1ex] adds vertical space
%\hline %inserts single line
%\end{tabular} \\
%Weibull(0.6,0.665)  
%& \hspace{1cm} (ii) Pareto(1.2,5)
\end{tabular}
 % is used to refer this table in the text
\caption{\small Parameters of the hyperexponential distribution fitted to Weibull($0.6$,$0.665$) (taken from Tables 3 of \cite{Feldmann_1998}).} \label{table:fitted}
\end{table}

Table \ref{table:fitted} shows the parameters of the
hyperexponential distribution fitted  to Weibull(0.6,0.665) when $m=6$.  As can be seen in Figure 4  of \cite{Feldmann_1998}, this fitting is very accurate.  We consider the \lev process in the form (\ref{levy_canonical}) where $Z$ is hyperexponential specified in Table \ref{table:fitted} as an approximation to Weibull(0.6,0.665).  We use various values of $\sigma$ with the common values of $\lambda = 1$, $\mu = 0.1$ and $q=0.03$.  The roots $\xi_{\cdot,q}$'s and  $\zeta_q$ are calculated
via the bisection method with error bound $1.0E-10$.

\begin{figure}[htbp]
\begin{center}
\begin{minipage}{1.0\textwidth}
\centering
\begin{tabular}{cc}
\includegraphics[scale=0.6]{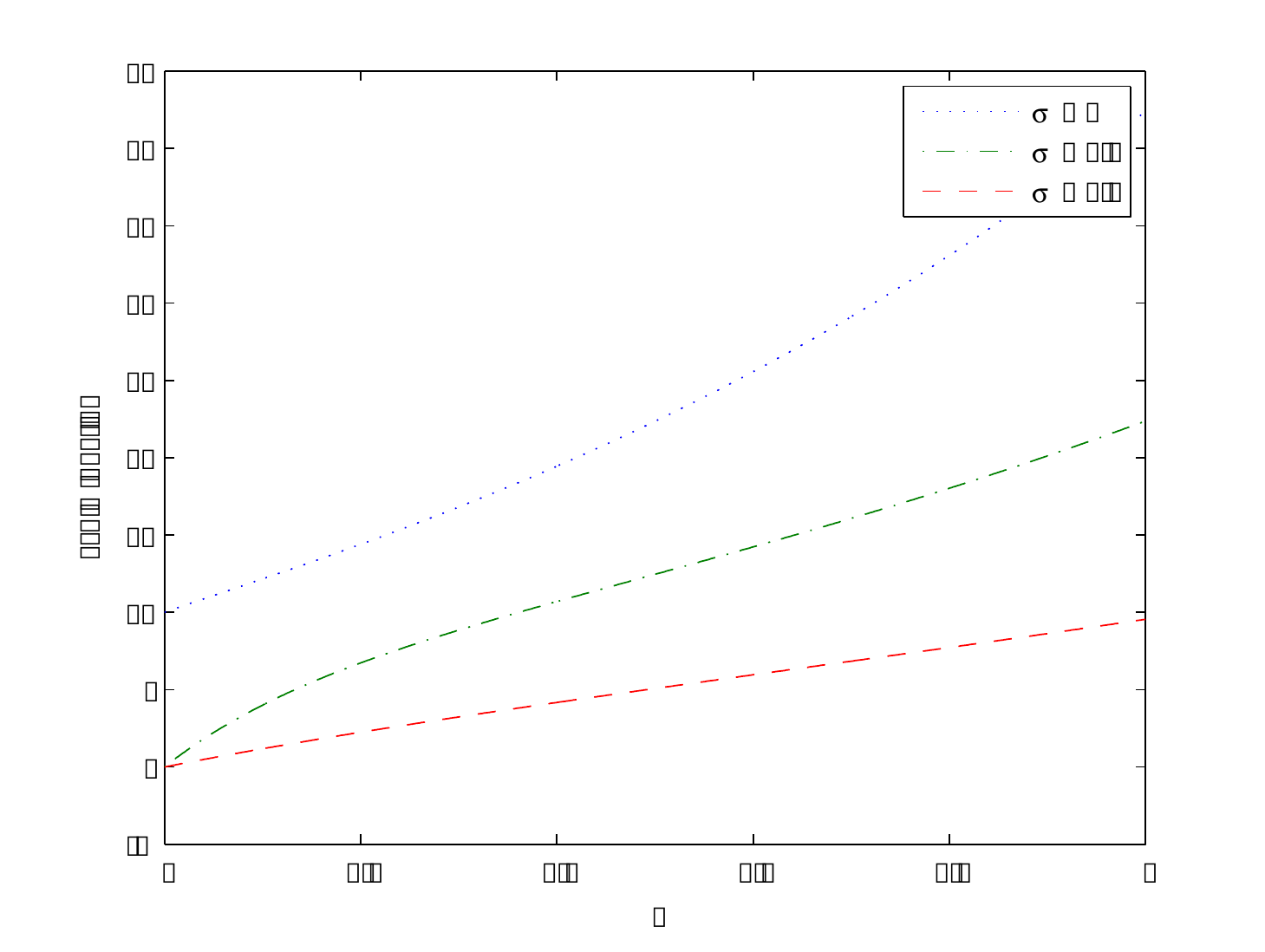}  & \includegraphics[scale=0.6]{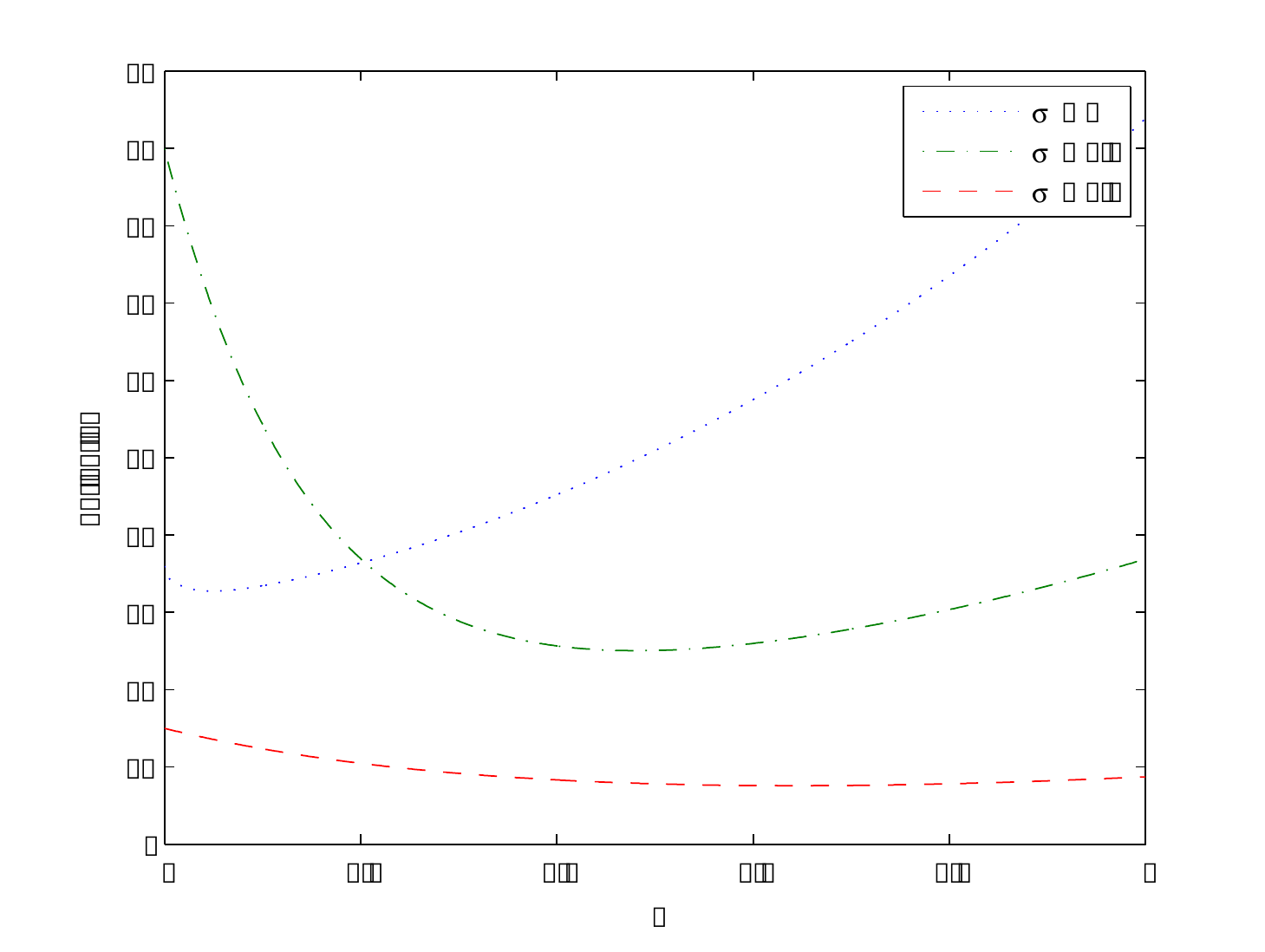} \\
scale function $W^{(q)}$ & derivative  $W^{(q)'}$ \vspace{0.5cm} 
\end{tabular}
\end{minipage}
\caption{Scale functions and their derivatives for the case with Weibull-distributed jumps.}
\label{fig:weibull_scale}
\end{center}
\end{figure}

\begin{figure}[htbp]
\begin{center}
\begin{minipage}{1.0\textwidth}
\centering
\includegraphics[scale=0.6]{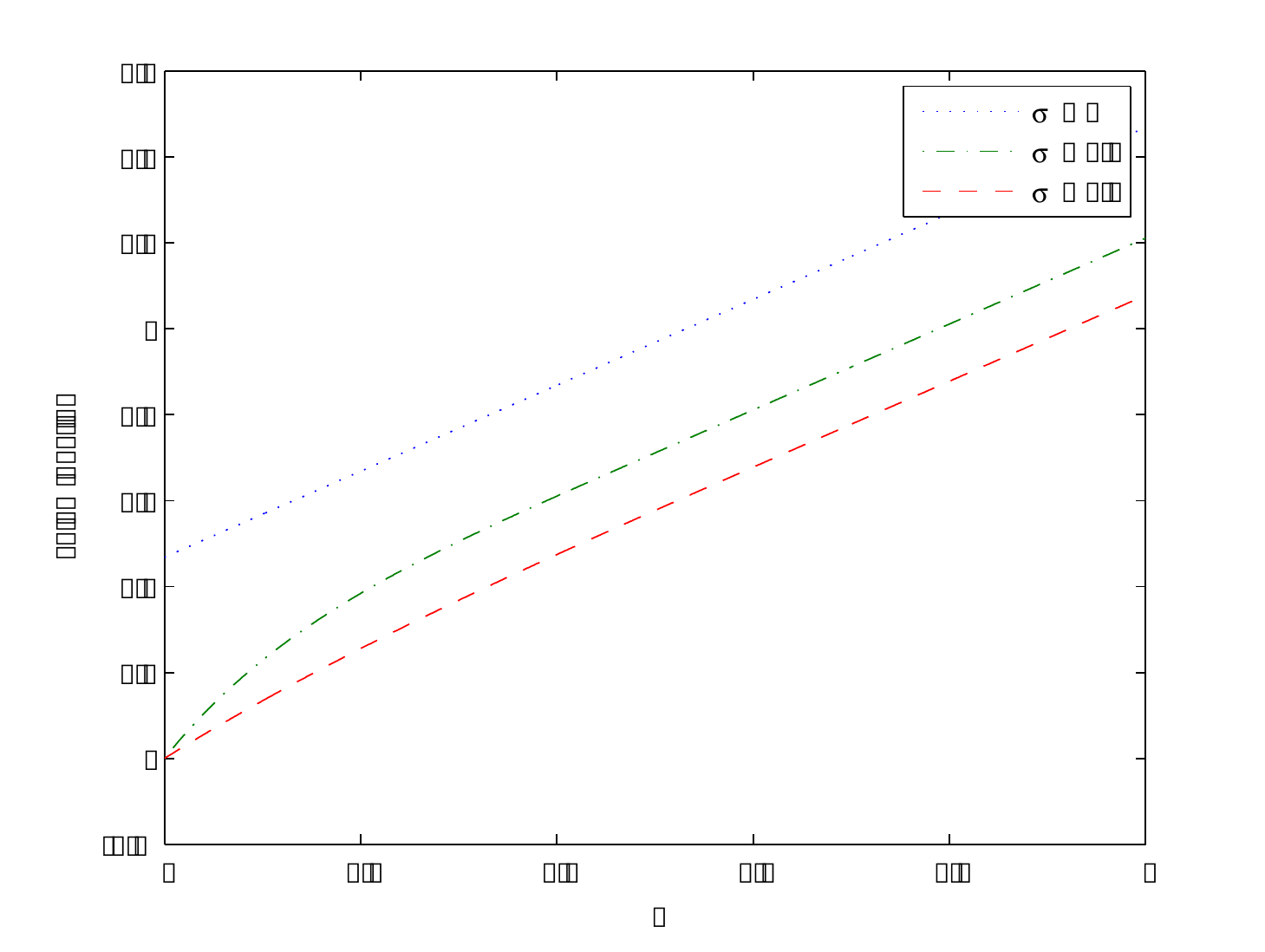} 
\end{minipage}
\caption{Value function $v_{\pi_{a^*}}$of the classical dividend problem.}
\label{fig:weibull_value_function}
\end{center}
\end{figure}

Figure \ref{fig:weibull_scale} shows the scale function $W^{(q)}$ and its derivative $W^{(q)'}$.  The optimal barrier levels that minimize  $W^{(q)'}$ are given by $a^*= 0.05,0.481$ and $0.643$ for the cases $\sigma=0,0.2$ and $0.4$, respectively.  The results are consistent with Lemma \ref{lemma_zero}; with the existence of a diffusion component, the scale function is forced to converge to $0$ as $x$ goes to $0$.  Using these barrier levels, optimal value functions for the classical dividend problem can be computed by Theorem \ref{theorem_optimality}.  Figure \ref{fig:weibull_value_function} shows the value functions $v_{\pi_{a^*}}$.   Notice that they are monotonically decreasing in $\sigma$.

\begin{figure}[htbp]
\begin{center}
\begin{minipage}{1.0\textwidth}
\centering
\begin{tabular}{cc}
\includegraphics[scale=0.6]{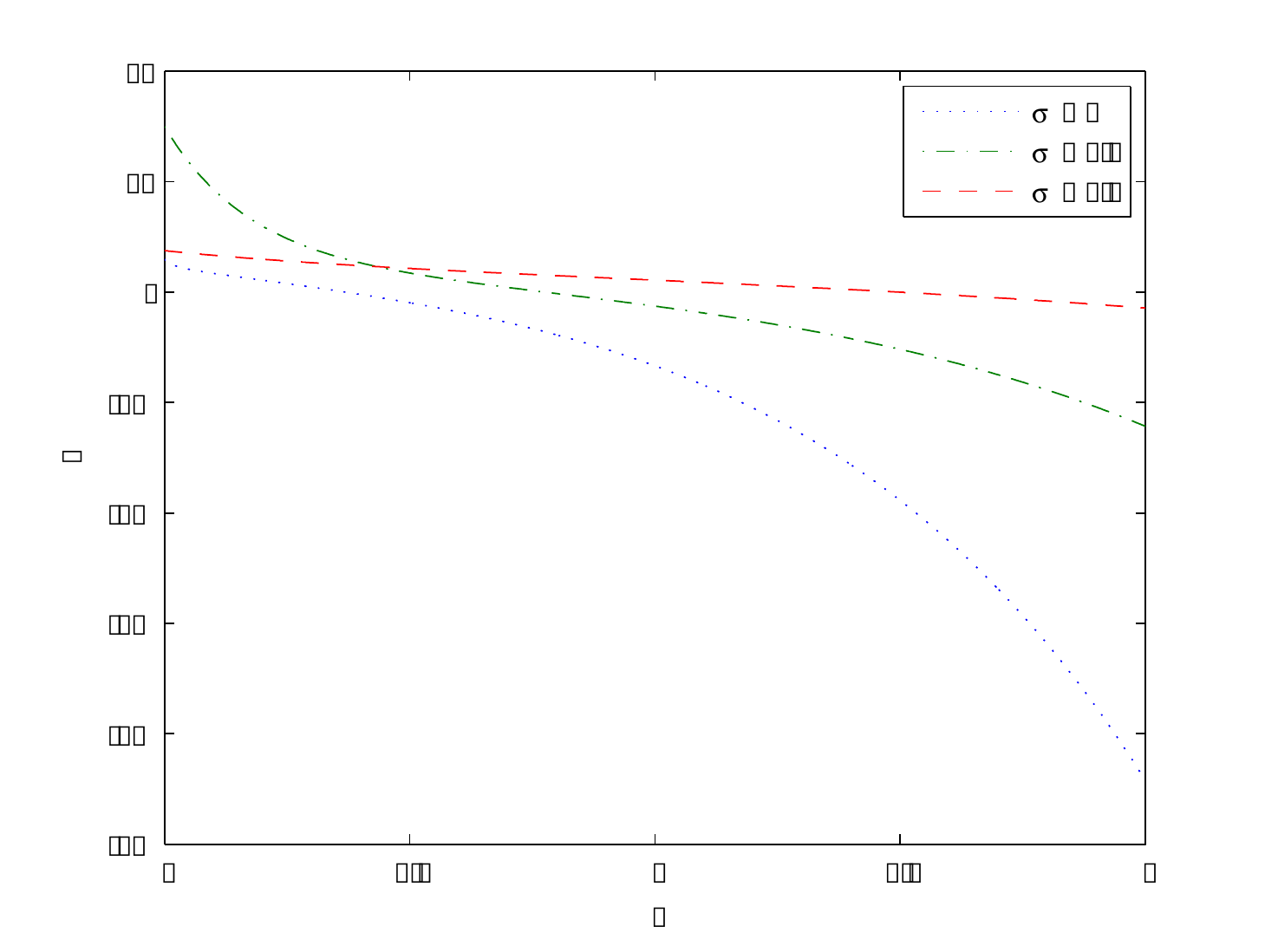}  & \includegraphics[scale=0.6]{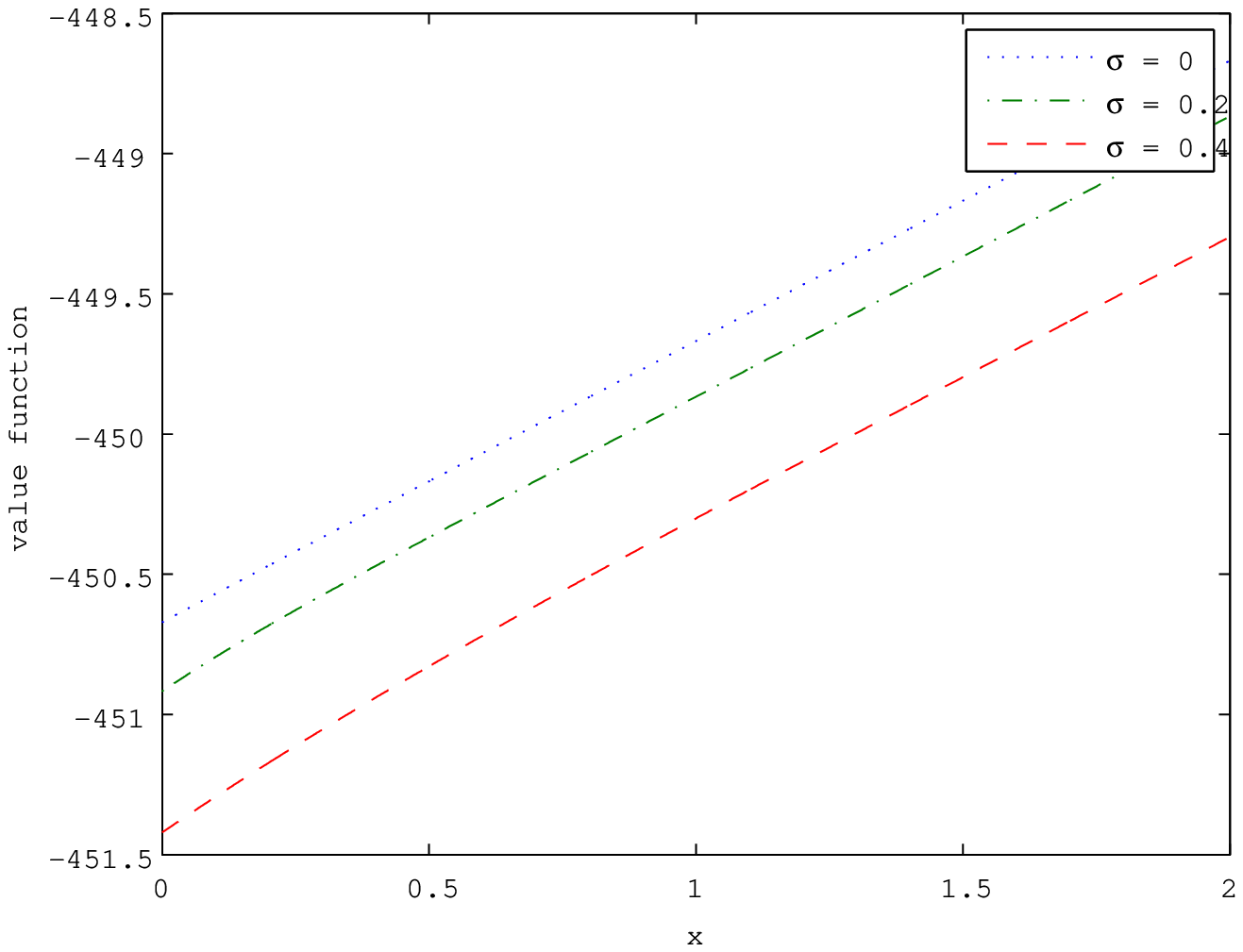} \\
$G(x)$  & value function $\overline{v}_{d^*}$ \vspace{0.5cm} 
\end{tabular}
\end{minipage}
\caption{Bail-out problem: $G$ in (\ref{optimal_barrier_bail_out}) and the value function  $\overline{v}_{d^*}$ when $\varphi = 1.3$.  Optimal barriers are $0.38$, $0.775$ and $1.495$ for $\sigma = 0,0.2$ and $0.4$, respectively.}
\label{fig:bail_out}
\end{center}
\end{figure}

\begin{figure}[htbp]
\begin{center}
\begin{minipage}{1.0\textwidth}
\centering
\begin{tabular}{cc}
\includegraphics[scale=0.6]{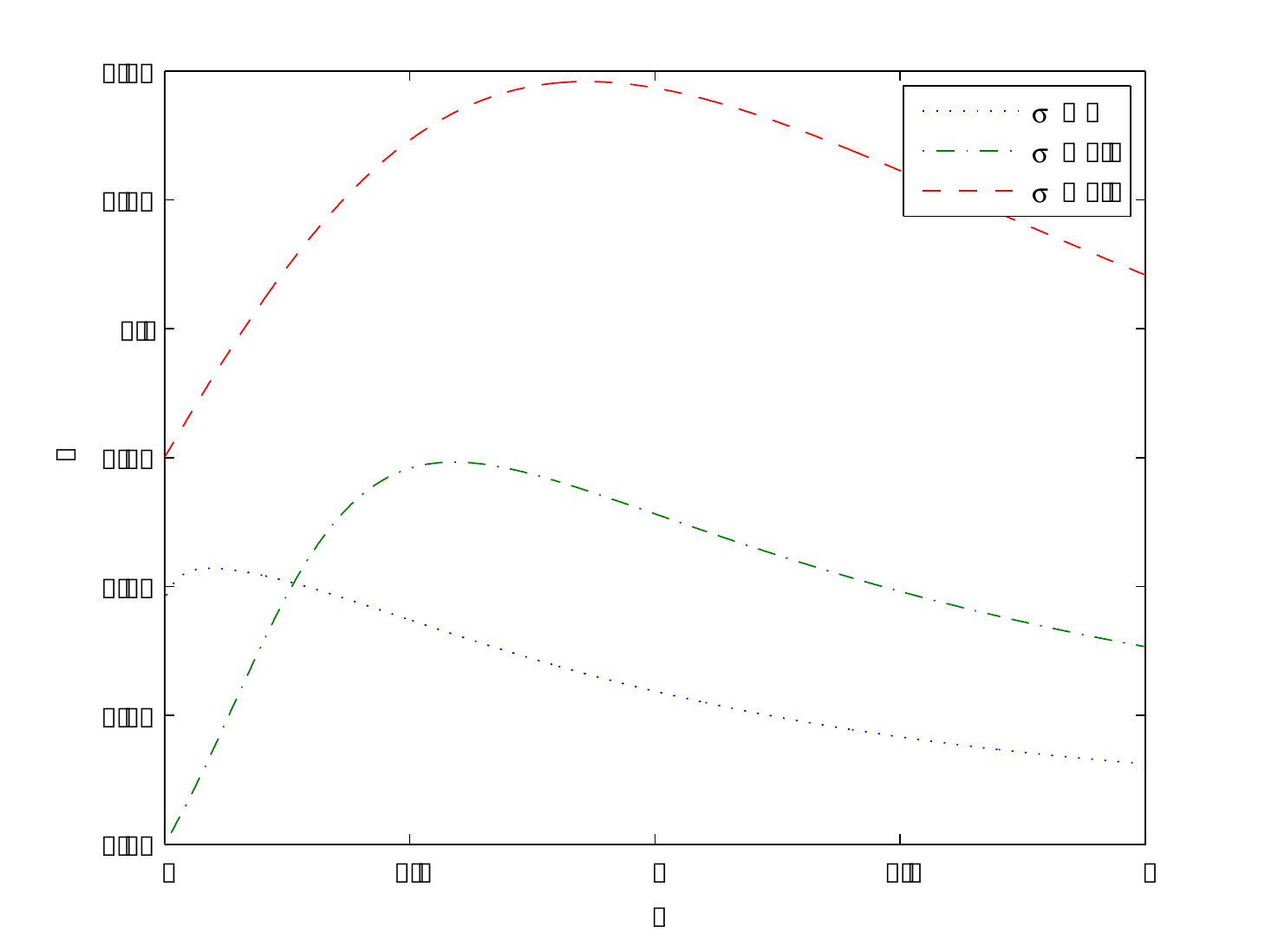}  & \includegraphics[scale=0.6]{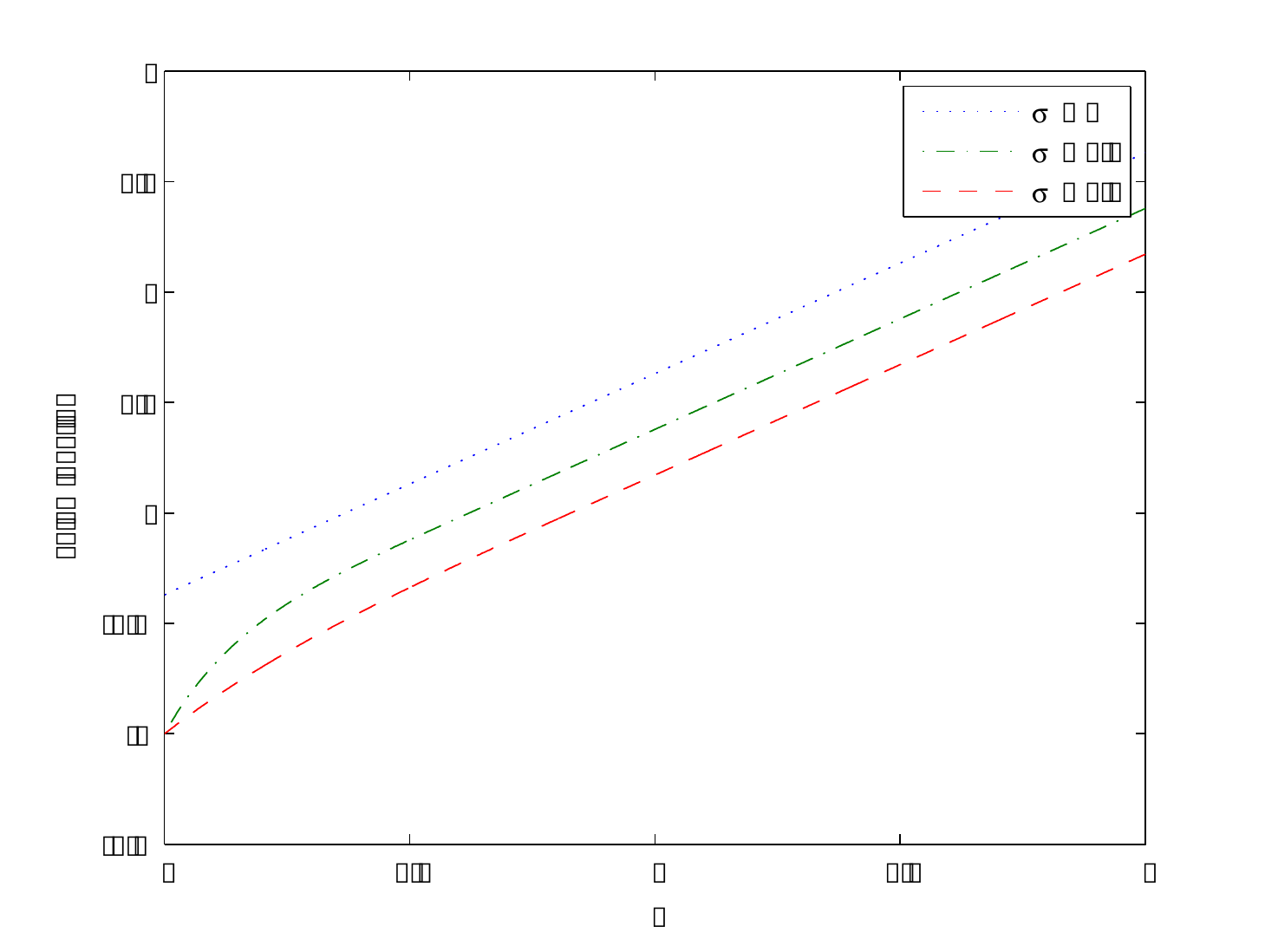} \\
 $F(x)$ when (i) $S=-1,K=0$ & value function $\widetilde{v}_{b^*}$ when (i) $S=-1,K=0$\\
\includegraphics[scale=0.6]{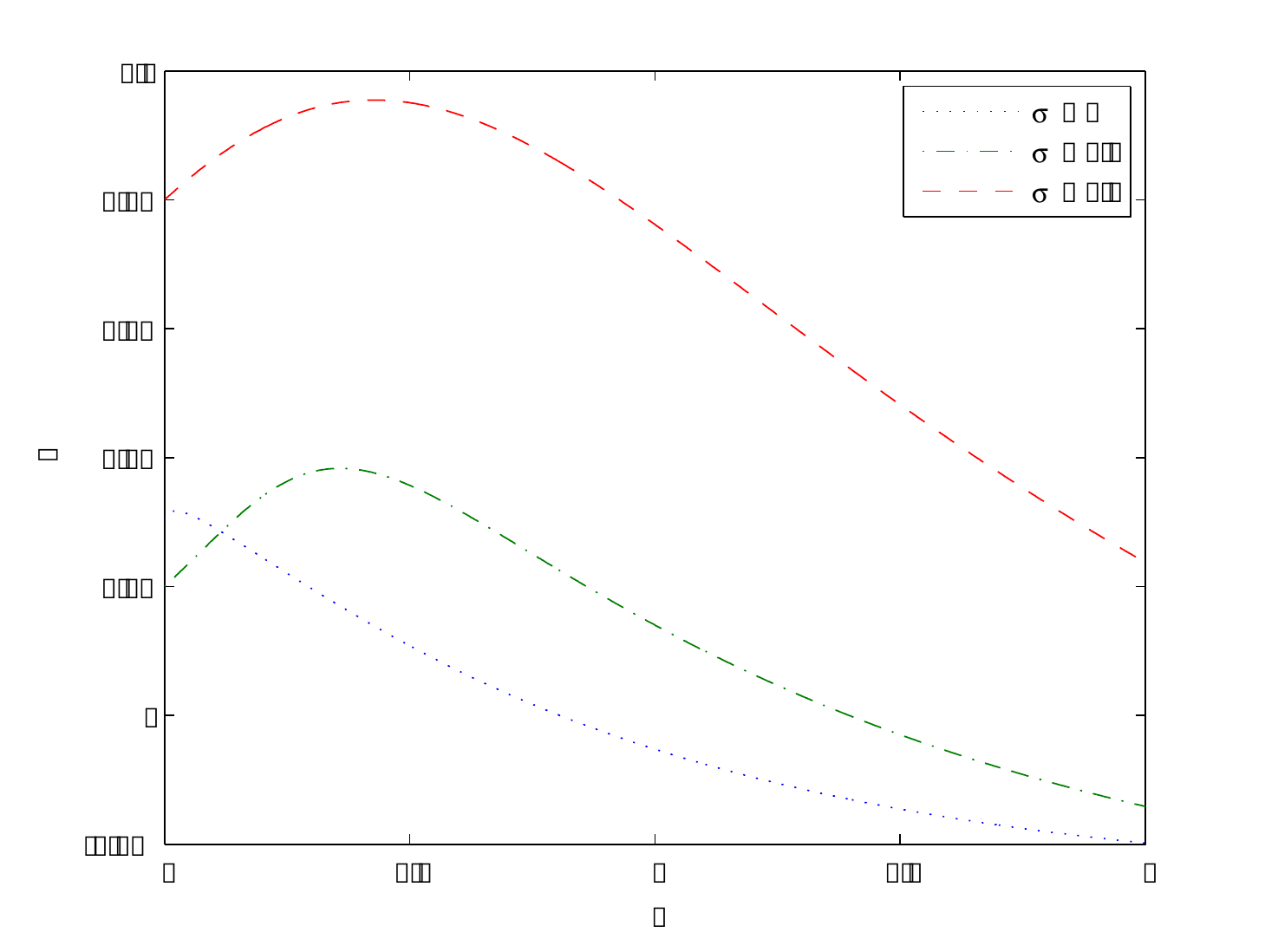}  & \includegraphics[scale=0.6]{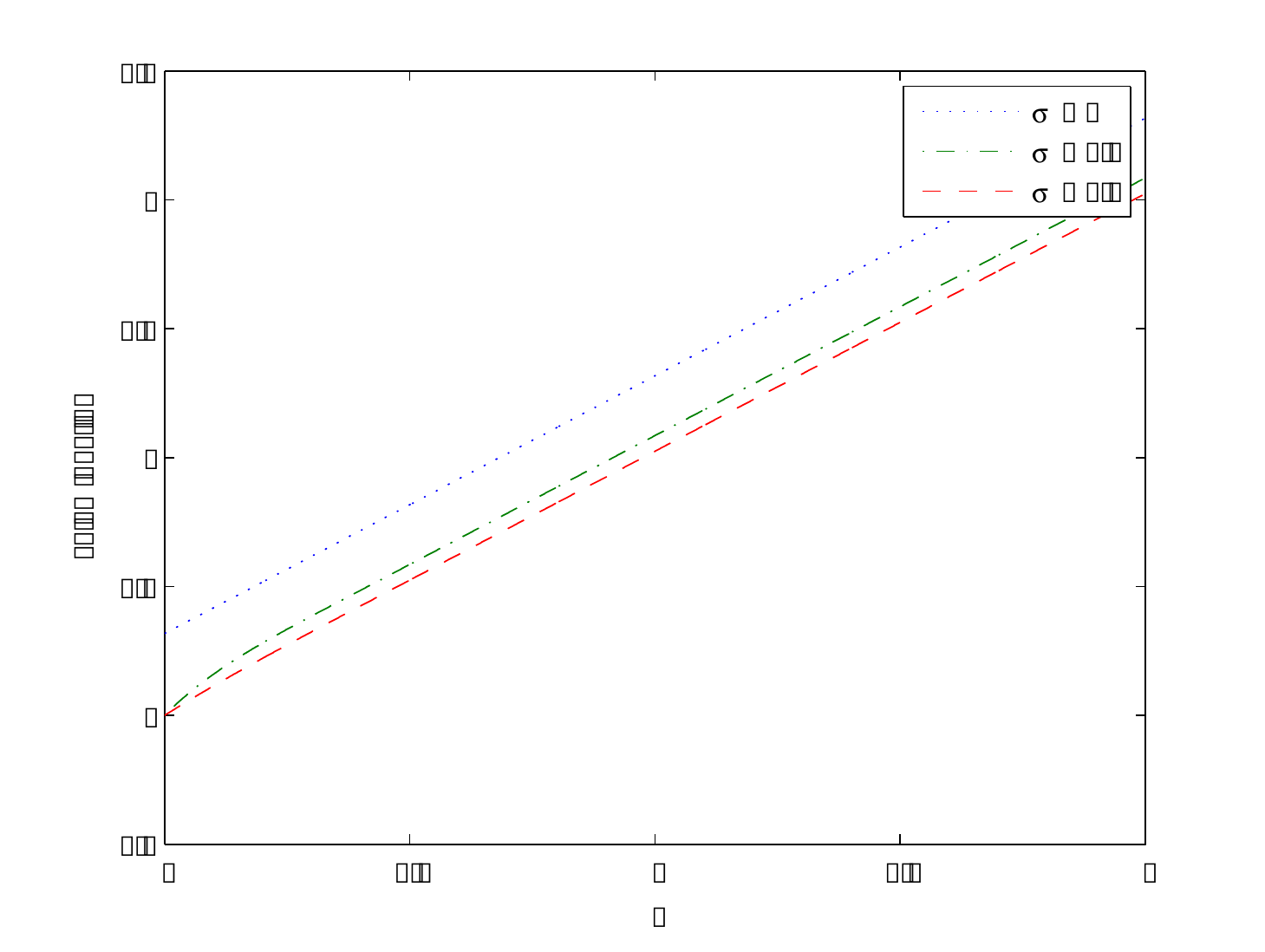} \\
 $F(x)$ when (ii) $S=1,K=0$ & value function $\widetilde{v}_{b^*}$ when (ii) $S=1,K=0$\\ \vspace{0.5cm} 
\end{tabular}
\end{minipage}
\caption{Extension with terminal values at ruin: $F$ in (\ref{F}) and the value function $\widetilde{v}_{b^*}$ when (i) $S=-1,K=0$ and (ii) $S=1,K=0$. The optimal barriers are (i) $0.0628$, $0.0793$ and $0.1384$ and (ii) $0.0317$, $0.0383$ and $0.0955$, respectively, for $\sigma = 0,0.2$ and $0.4$.}
\label{fig:weibull_terminal}
\end{center}
\end{figure}

\begin{figure}[htbp]
\begin{center}
\begin{minipage}{1.0\textwidth}
\centering
\begin{tabular}{cc}
\includegraphics[scale=0.6]{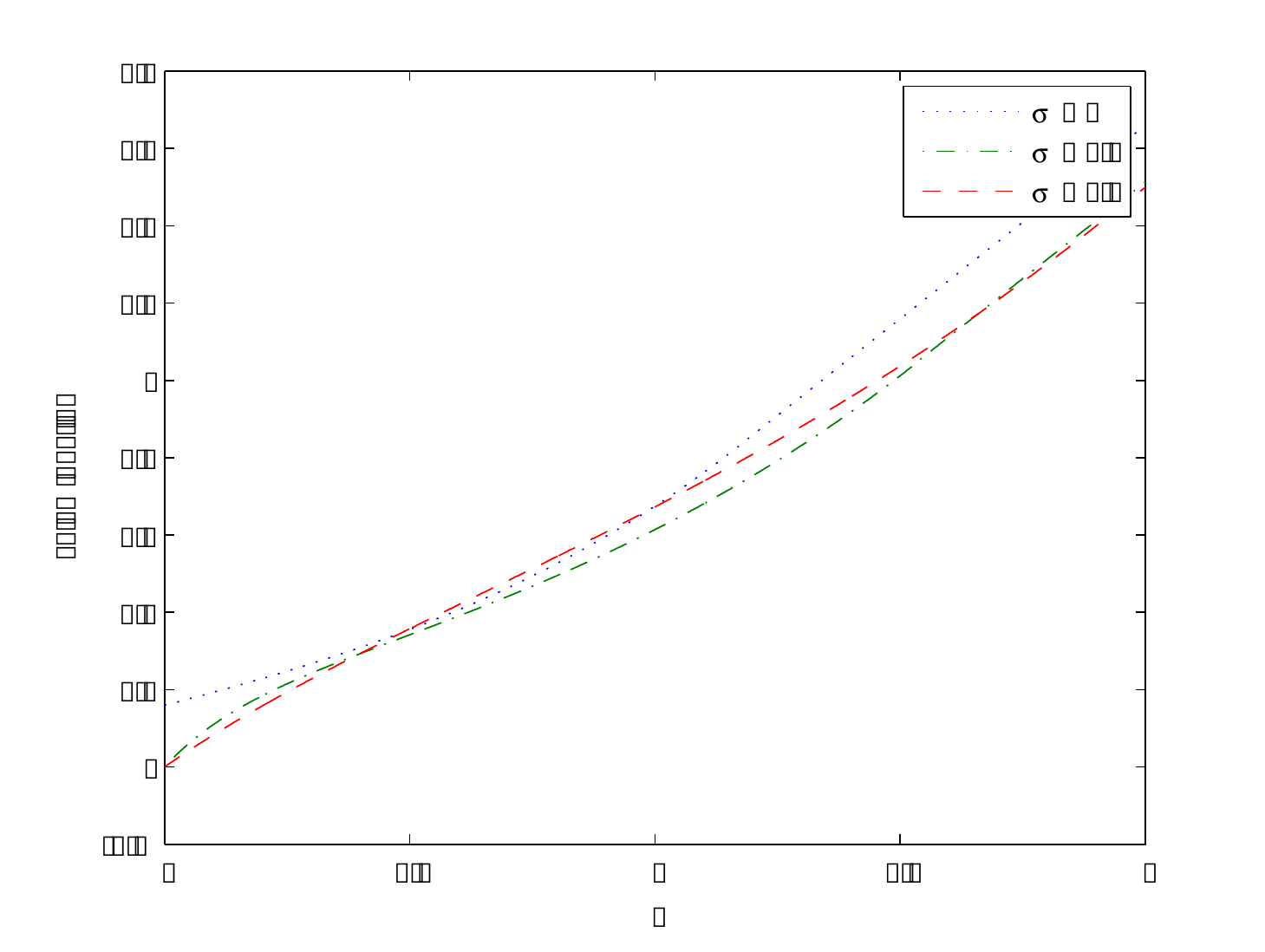}  & \includegraphics[scale=0.6]{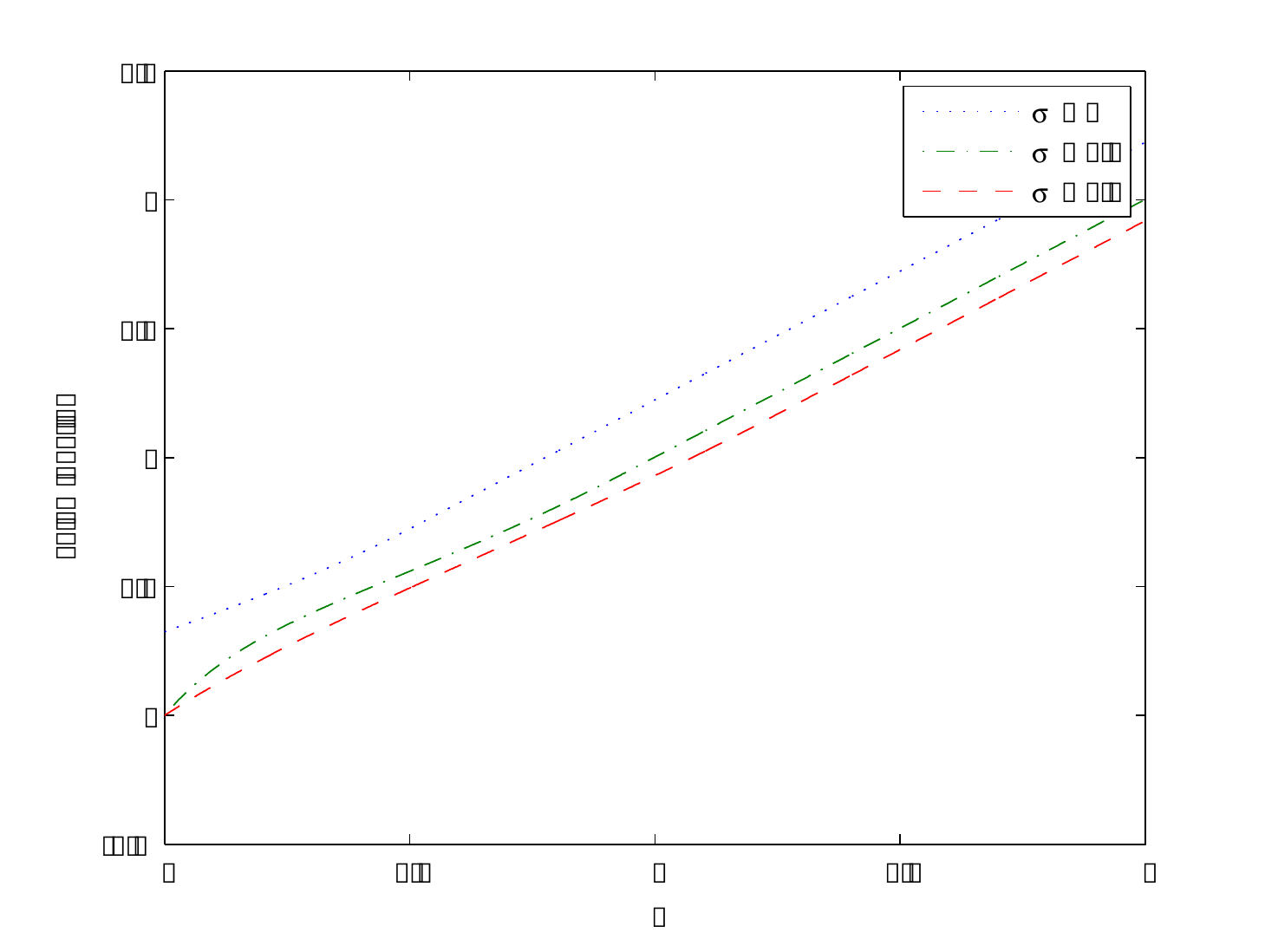} \\
value function $v_{c_2^*}$ when (i)  $\beta = 0.5$& value function $v_{c_2^*}$ when (ii) $\beta=0.1$
\end{tabular}
\end{minipage}
\caption{Extension with transaction costs: value function when (i) $\beta = 0.5$ and (ii) $\beta=0.1$. The optimal impulse controls $(c_1^*,c_2^*)$ are (i) $(0,1.173)$, $(0.069,1.527)$ and $(0,1.885)$ and (ii) $(0,0.05)$, $(0.222,0.481)$ and $(0.197,0.643)$  for $\sigma = 0,0.2$ and $0.4$, respectively.}
\label{fig:weibull_transaction}
\end{center}
\end{figure}

We now consider the extensions described in the end of Section \ref{section_dividend_problem}.  Here we use the same parameters as in the results above. Figure \ref{fig:bail_out} shows the results on the bail-out problem \citep{Avram_et_al_2007} when $\varphi = 1.3$. It plots $G$ in (\ref{optimal_barrier_bail_out}) as well as the value function $\overline{v}_{d^*}$ in (\ref{def_v_bar}).  Here the optimal barrier level is obtained by computing the unique level $d^*$ that satisfies $G(d^*)=0$.  Figure \ref{fig:weibull_terminal} shows the results on the extension with terminal values at ruin \citep{Loeffen_2010} with the plots of $F$ in (\ref{F}) and the value function $\widetilde{v}_{b^*}$  in \eqref{v_tilde}.  We consider the case with constant terminal value (i) $S=-1$ and $K=0$ and (ii) $S=1$ and $K=0$. The maximizer of $F$ becomes the barrier level $b^*$.  Figure \ref{fig:weibull_transaction} shows the value functions $v_{c_2^*}$ on the extension with transaction costs  \citep{Loeffen_2009_2} when (i) $\delta = 0.5$ and (ii) $\delta=0.1$.  In order to obtain the optimal impulse contol $(c_1^*,c_2^*)$, we use the technique discussed in  Section 4 of \cite{Loeffen_2009_2}.  Unlike the other results, the value functions are no longer monotone in $\sigma$ unless $\delta$ is sufficiently small.   However, as $\delta$ decreases to zero, $c_1^*$ converges to $c_2^*$ and the value function converges to that of the classical model as shown in Figure \ref{fig:weibull_value_function}.

\subsection{Numerical results on the $\beta$-class and CGMY process}
We now consider, as an example of meromorphic \lev processes, the $\beta$-class introduced by \cite{Kuznetsov_2010_2}. The following definition is due to \cite{Kuznetsov_2010_2}, Definition 4.
\begin{definition}
A spectrally negative \lev process is said to be in the $\beta$-class if its \lev measure is in the form
\begin{align}
\nu(\diff x) = c \frac {e^{\alpha \beta x}} {(1-e^{\beta x})^\lambda} 1_{\{ x < 0 \}} \diff x, \quad x \in \mathbb{R}, \label{levy_measure_beta}
\end{align}
for some $\alpha > 0$, $\beta > 0$, $c \geq 0$ and $\lambda \in (0,3)$.  It is equivalent to saying that its Laplace exponent is
\begin{align*}
\psi(z) = \hat{\mu} z + \frac 1 2 \sigma^2 z^2 + \frac c \beta \left\{ B(\alpha + \frac z \beta, 1 - \lambda) - B(\alpha, 1 - \lambda) \right\}
\end{align*}
where $B$ is the beta function $B(x,y):=\Gamma(x)\Gamma(y)/\Gamma(x+y)$. 
\end{definition}
The special case $\sigma = 0$ and $\beta = 1$ reduces to the class of \emph{Lamperti-stable} processes, which are obtained by the Lamperti transformation (\cite{Lamperti_1972}) from the stable processes conditioned to stay positive; see \cite{Bertoin_Yor_2001} and \cite{Caballero_2008} and references therein.  For  the scale function of a related process, see \cite{Kyprianou_2008}.

It can be also seen that this is a ``discrete-version" of the (spectrally negative) \emph{CGMY} process, whose \lev measure is given by
\begin{align}
\nu(\diff x) = c \frac {e^{\alpha x}} {|x|^\lambda} 1_{\{ x < 0 \}} \diff x, \quad x \in \R. \label{levy_measure_tempered}
\end{align}
Indeed, if we set  $c = \widetilde{c} \beta^\lambda$ and $\alpha = \widetilde{\alpha} \beta^{-1}$ in (\ref{levy_measure_beta}), we have
\begin{align}
c \frac {e^{\alpha \beta x}} {(1-e^{\beta x})^\lambda} 1_{\{ x < 0 \}} \xrightarrow{\beta \downarrow 0} \widetilde{c} \frac {e^{\widetilde{\alpha} x}} {|x|^\lambda} 1_{\{ x < 0 \}}, \quad x \in \mathbb{R}. \label{convergence_cgmy}
\end{align}
%The \lev measure (\ref{levy_measure_tempered}) is able to model a number of processes.  
%For example, if $\lambda < 1$, we have
%\begin{align*}
%\nu(\mathbb{R}) = c (\alpha)^{-(1+\lambda)} \Gamma(1-\lambda) < \infty
%\end{align*}
%and the corresponding pure jump process becomes a compound Poisson process.  If $\lambda = 1$, it becomes the \emph{gamma process} ($\nu(\mathbb{R}) = \infty$).  If $\lambda = 2$, it becomes the \emph{Cauchy process}. 
See \cite{AsmussenMadanPistorius07} for approximation of (double-sided) CGMY processes using hyperexponential distributions.

\begin{figure}[htbp]
\begin{center}
\begin{minipage}{1.0\textwidth}
\centering
\begin{tabular}{cc}
%(a) $m=100$ & (b) $m=1000$\\
\includegraphics[scale=0.5]{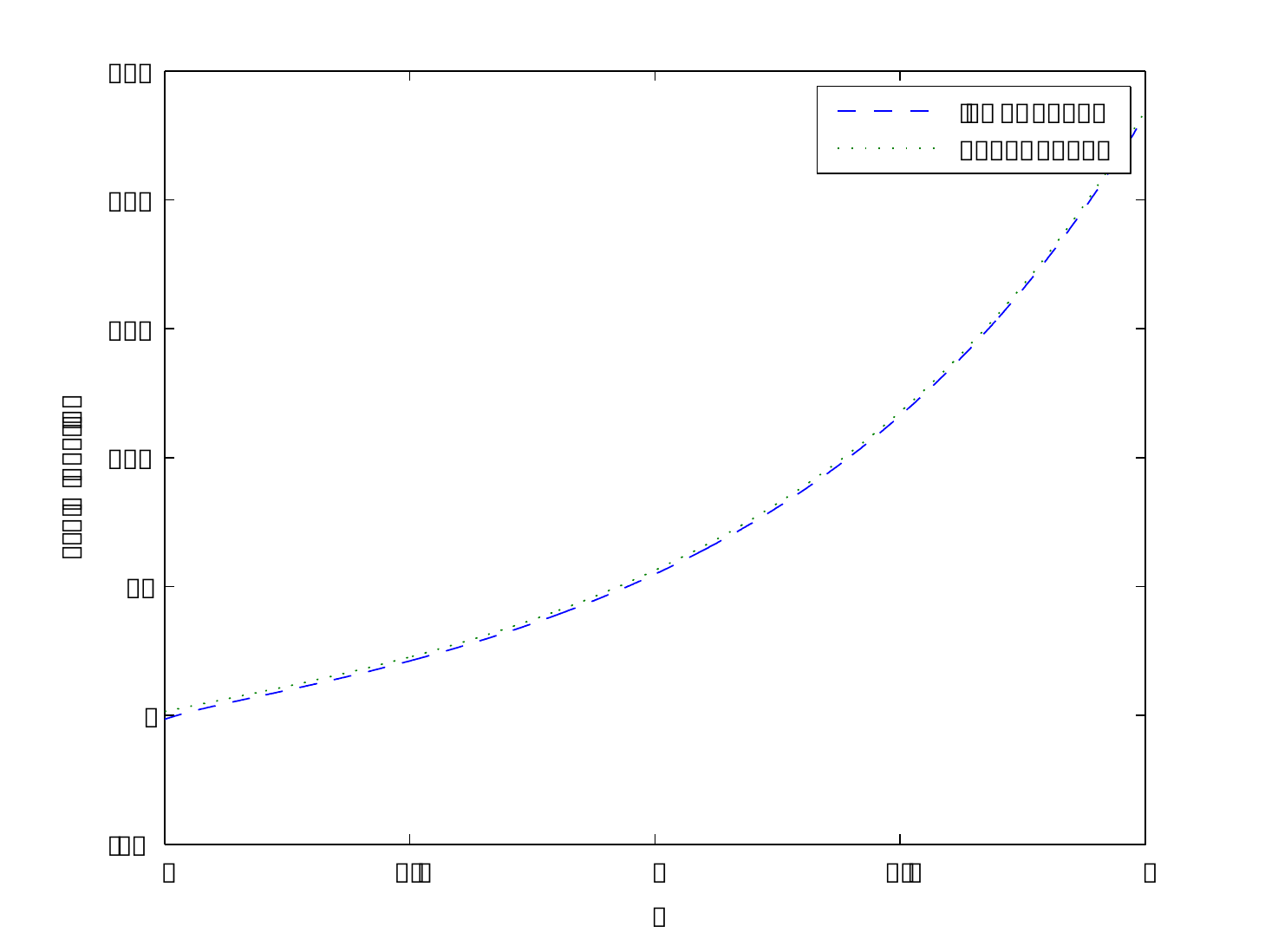}  & \includegraphics[scale=0.5]{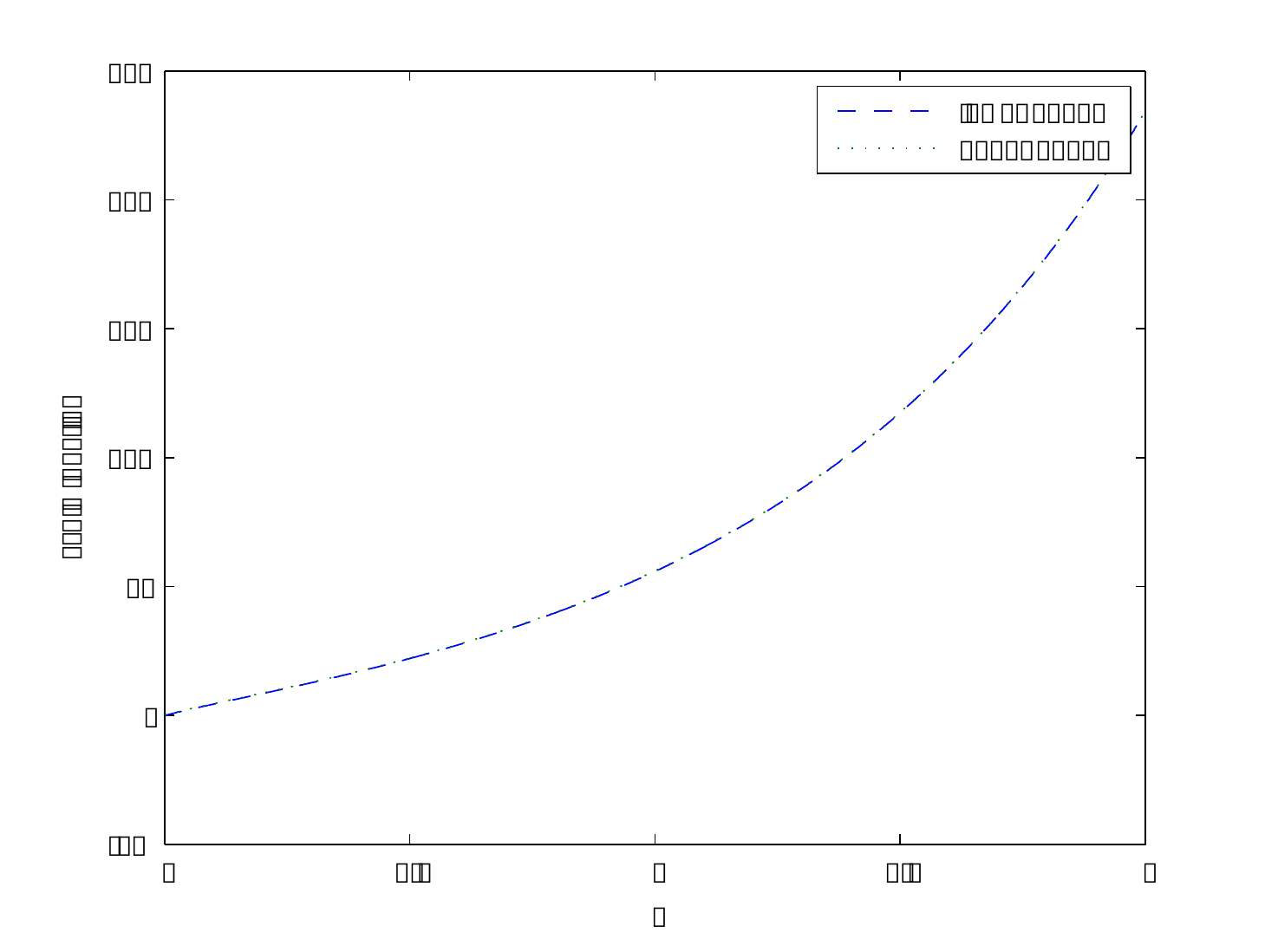} \\
Bounds on $W^{(q)}(x)$ when $m=15$ & Bounds on $W^{(q)}(x)$ when $m=150$ \vspace{0.5cm} \\
\includegraphics[scale=0.5]{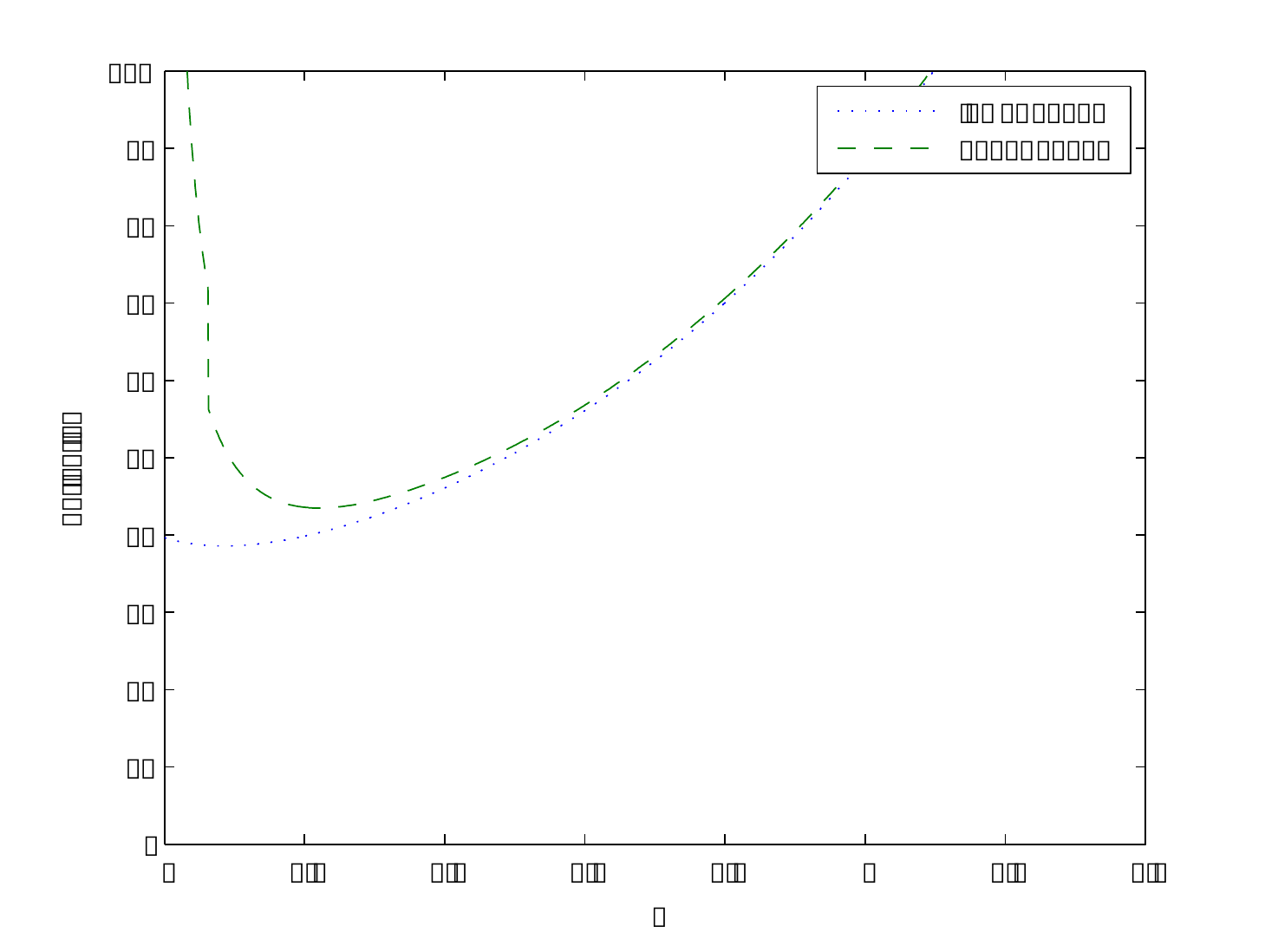}  & \includegraphics[scale=0.5]{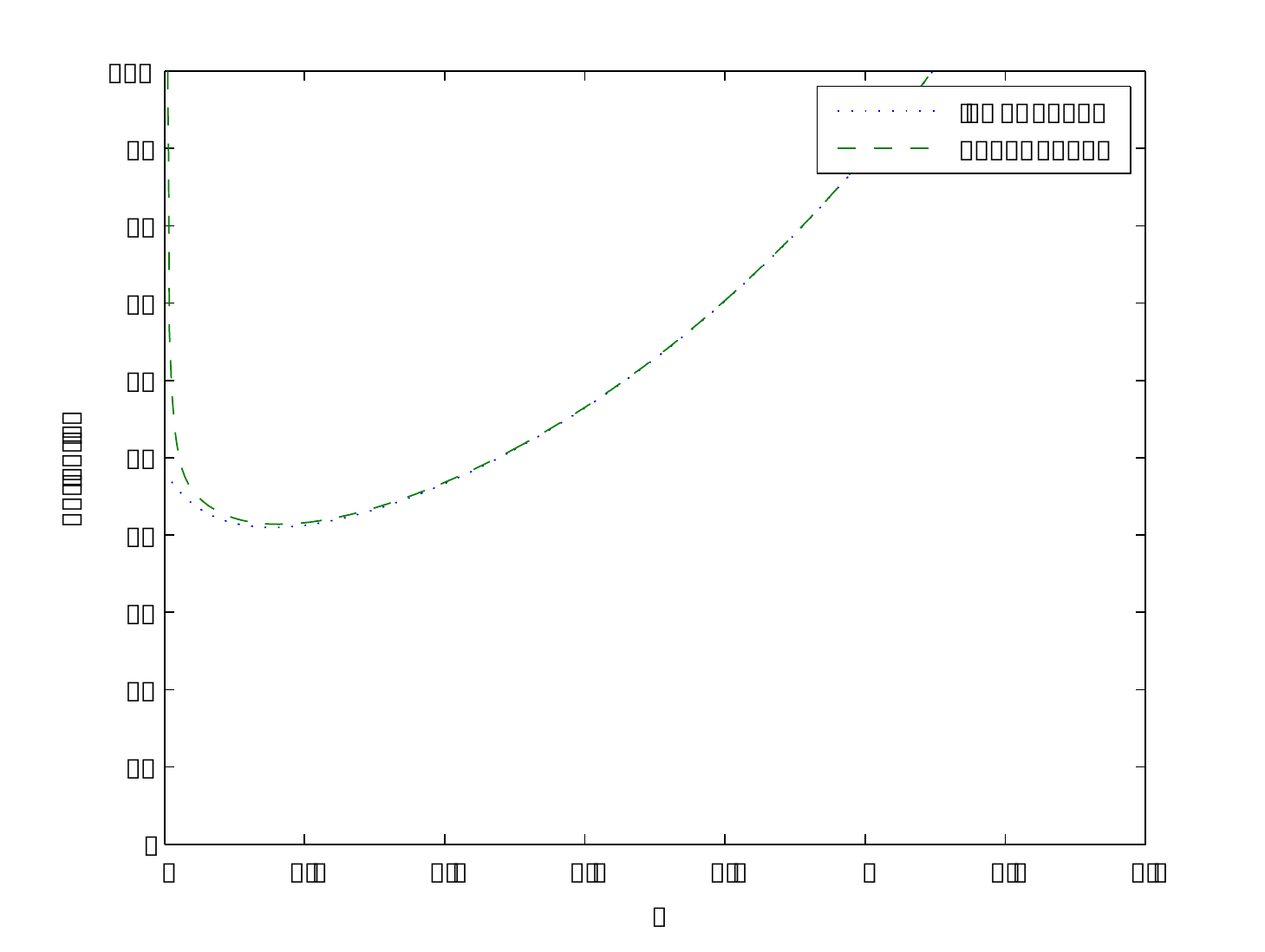} \\
Bounds on $W^{(q)'}(x)$ when $m=15$ & Bounds on $W^{(q)'}(x)$ when $m=150$ \vspace{0.5cm} \\
\includegraphics[scale=0.5]{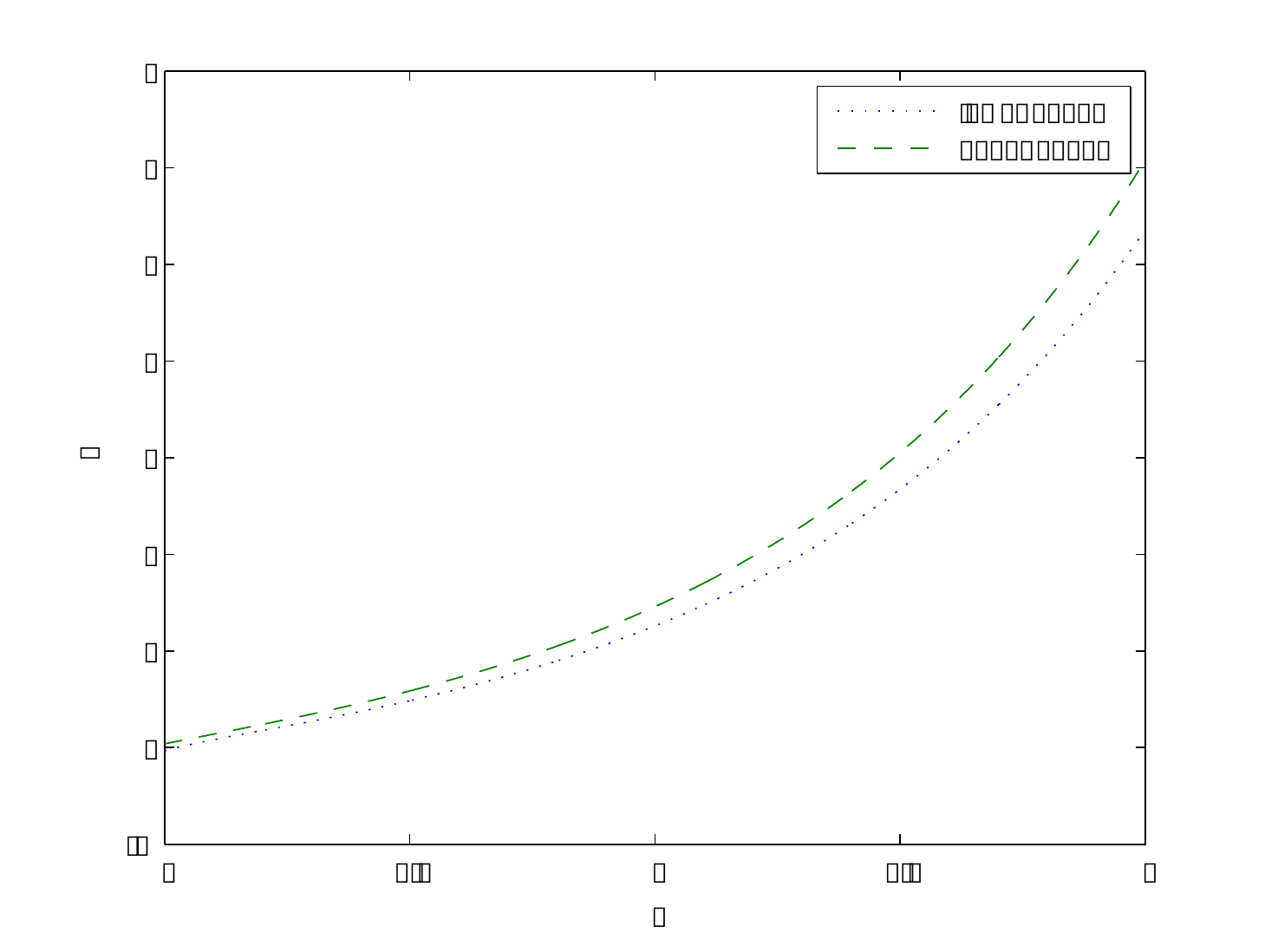}  & \includegraphics[scale=0.5]{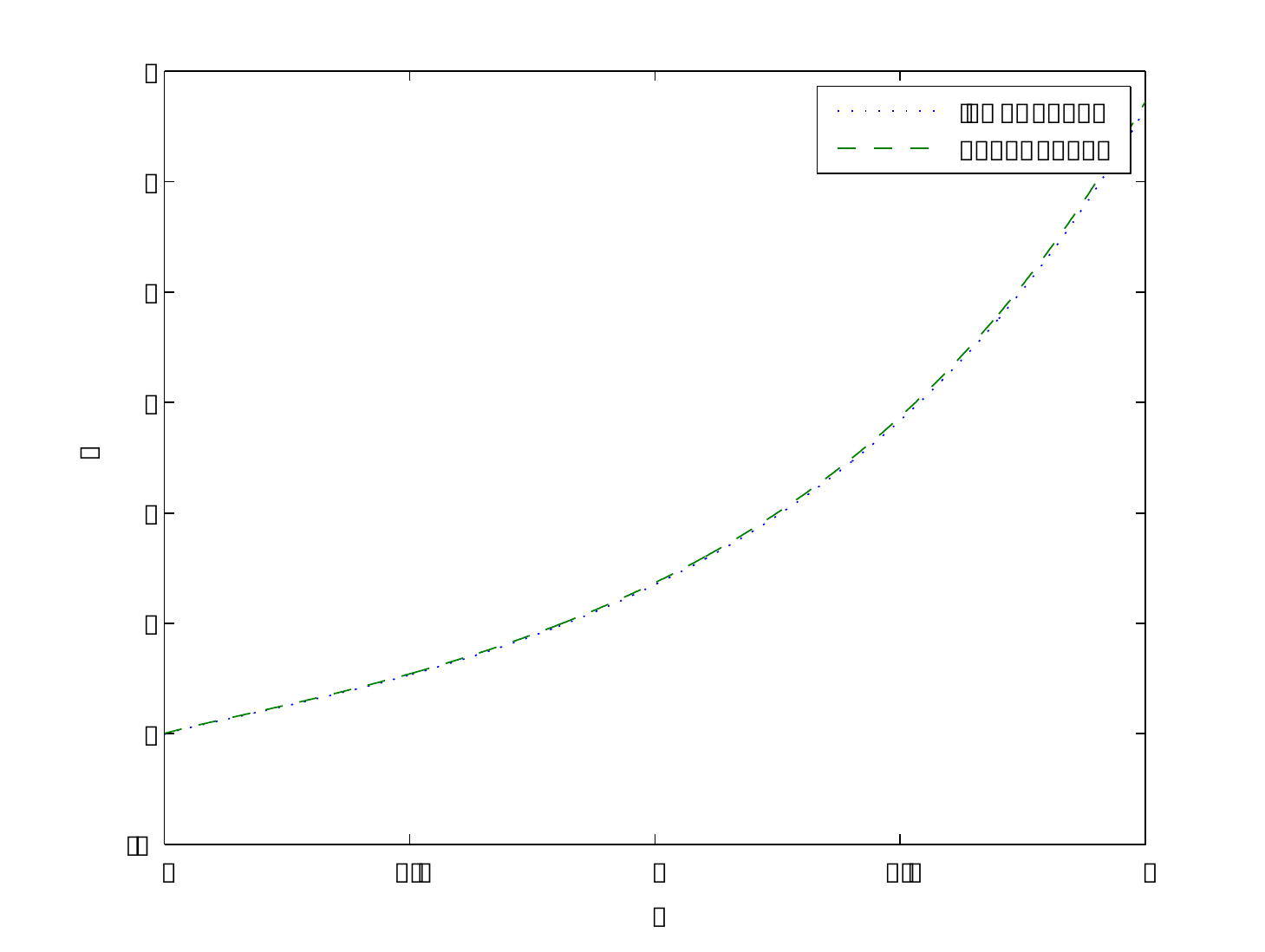} \\
Bounds on $u(x)$ when $m=15$ & Bounds on $u(x)$ when $m=150$
\end{tabular}
\end{minipage}
\caption{Approximation of the scale function and its derivative for the $\beta$-class.}
\label{fig:1}
\end{center}
\end{figure}

We shall use the results in Section \ref{section_meromorphic} to obtain the bounds on the scale functions and the solutions to the classical optimal dividend problem.  Figure \ref{fig:1} shows the approximation results when $q = 0.03$, $\sigma = 0.2$, $\hat{\mu} = 0.1$, $\lambda = 1.5$, $\alpha=3$, $\beta=1$ and $c = 0.1$ in \eqref{levy_measure_beta}.  We plot, for $m=15$ and $m=150$, the upper and lower bounds on the scale function, its derivative and the function $u_{a^*}$ defined in \eqref{def_u}. As shown in the previous section, the difference between the upper and lower bounds indeed converges to zero.  

\begin{figure}[htbp]
\begin{center}
\begin{minipage}{1.0\textwidth}
\centering
\begin{tabular}{cc}
%scale functions & derivatives \\
\includegraphics[scale=0.6]{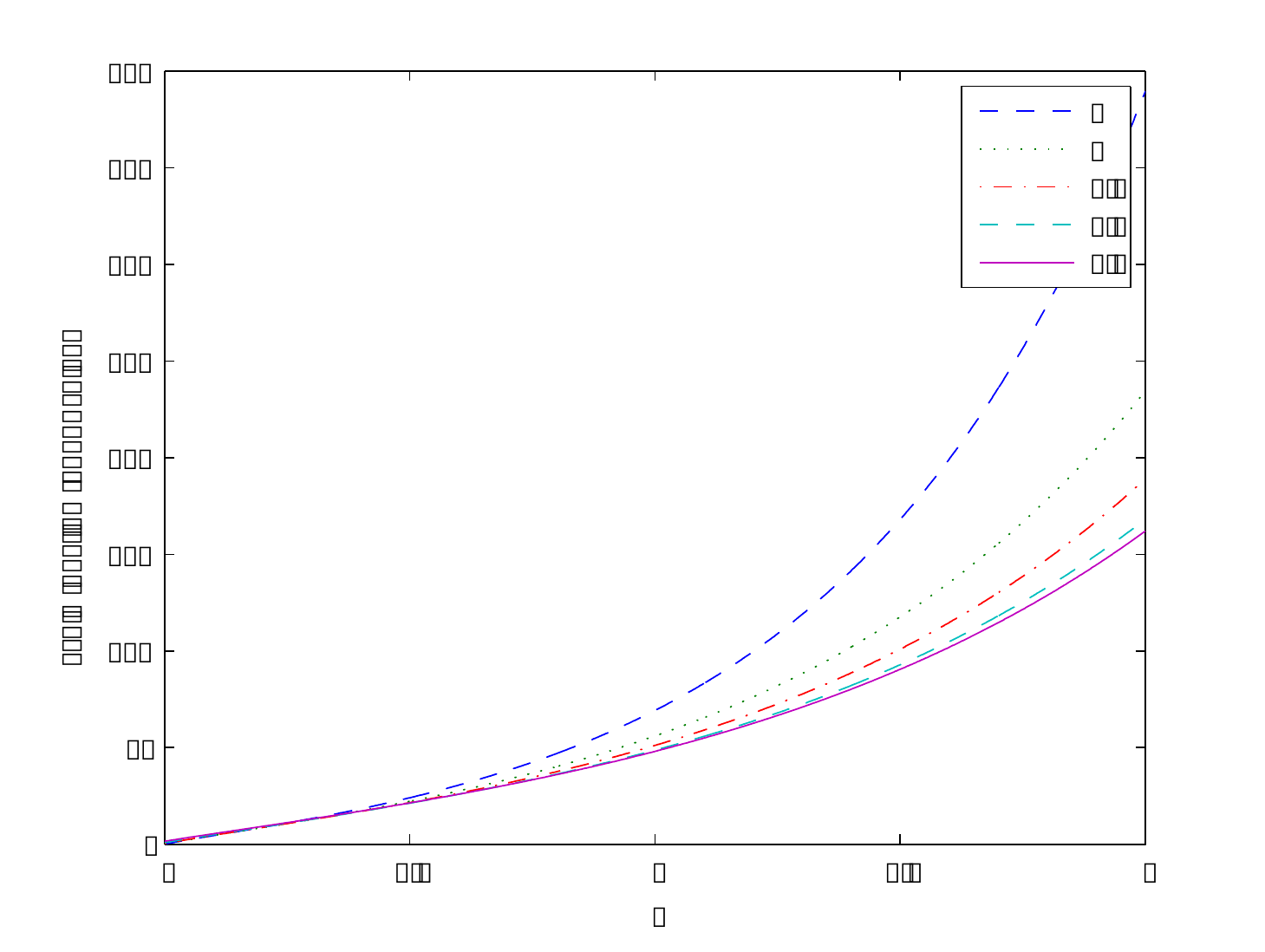}  & \includegraphics[scale=0.6]{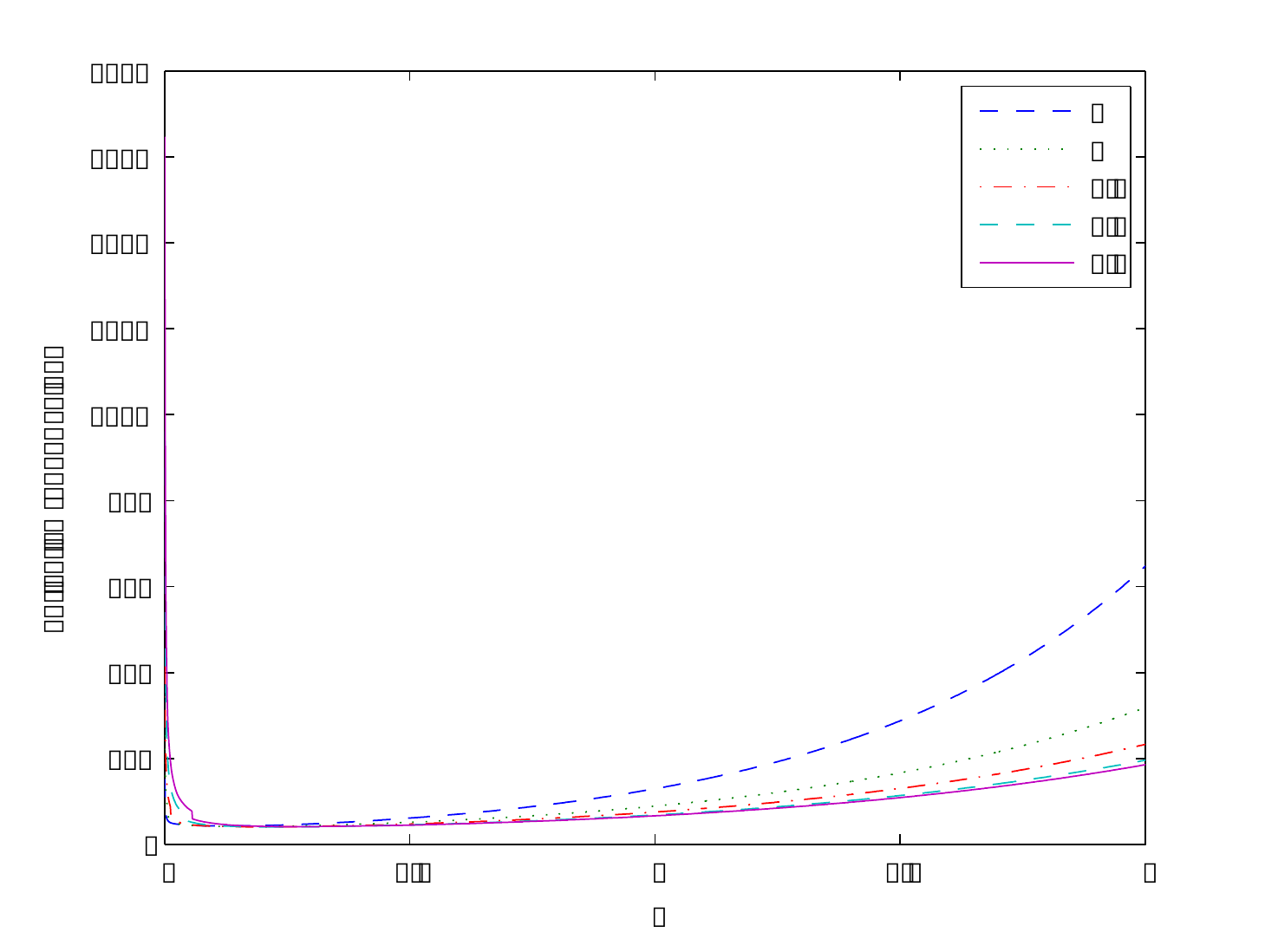} \\
upper bounds for the scale functions & upper bounds for the derivatives \vspace{0.5cm} \\
\includegraphics[scale=0.6]{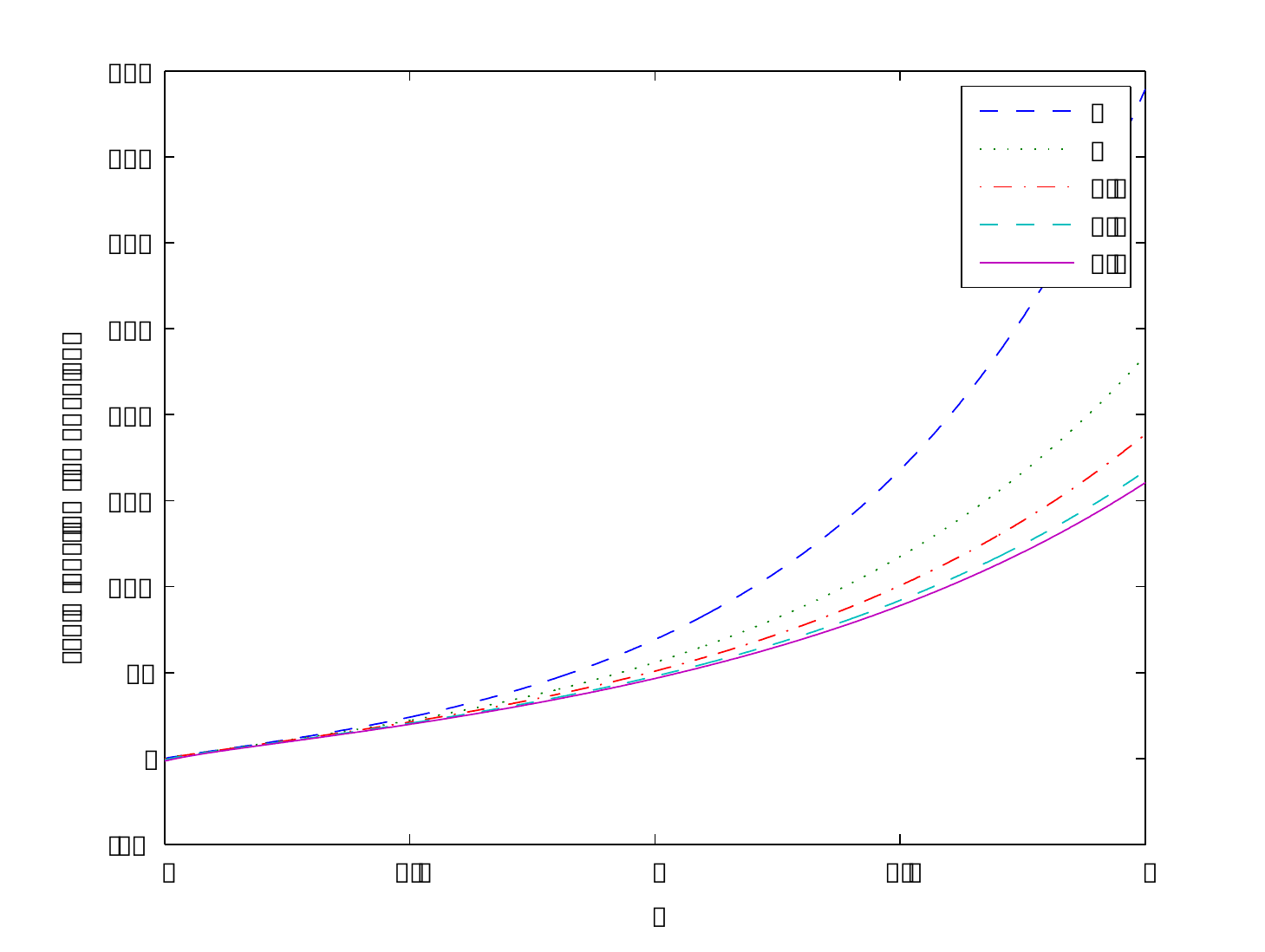}  & \includegraphics[scale=0.6]{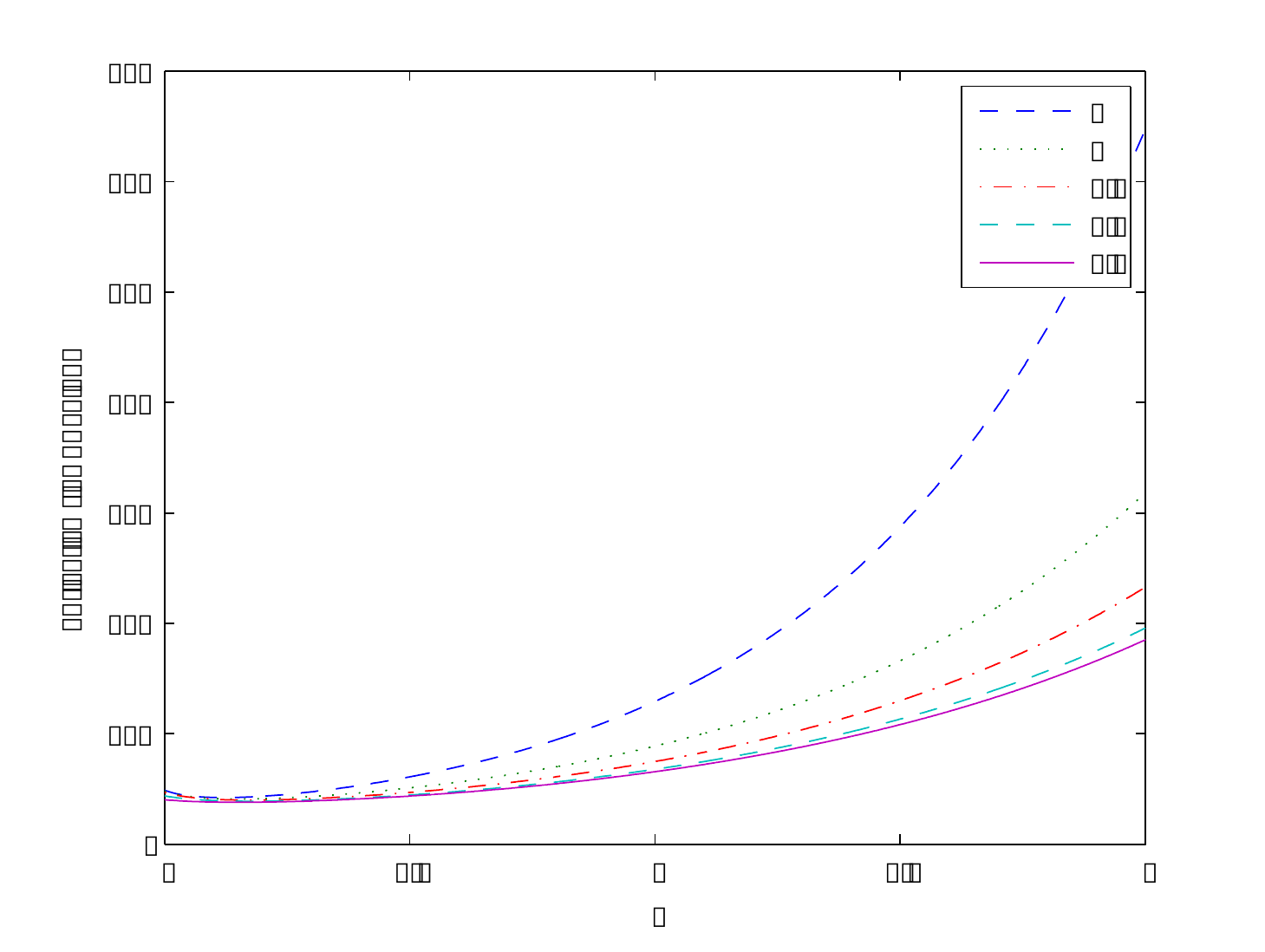} \\
lower bounds for the scale functions & lower bounds for the derivatives \vspace{0.5cm} \\
\end{tabular}
\end{minipage}
\caption{Convergence of scale functions to the CGMY model.}
\label{fig:3}
\end{center}
\end{figure}

\begin{figure}[htbp]
\includegraphics[scale=0.8]{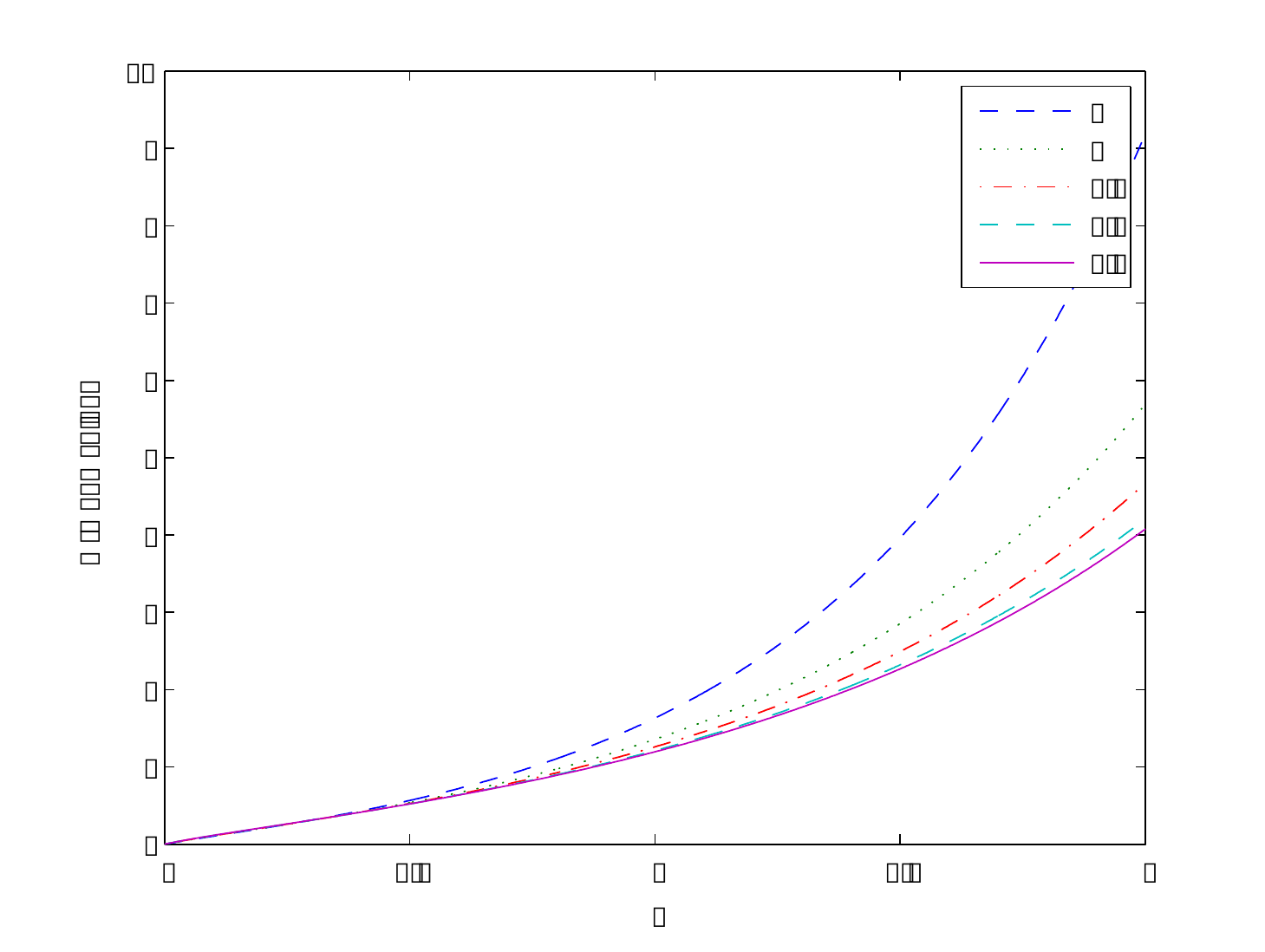}
\caption{Convergence of $u$ to the CGMY model. Here we use $M=150$. }
\label{fig:4}
\end{figure}

We now take $\beta$ in (\ref{convergence_cgmy}) to zero and see how the approximation for the CGMY process works.  Here we set $\tilde{\alpha}=3$ and $\tilde{c} = 0.1$ and use the same values as the above for the other parameters. Figure \ref{fig:3} shows the upper and lower bounds of scale function and its derivative for various values of $\beta$.  Figure \ref{fig:4} shows the mean value of the upper and lower bounds on the function $u$.   Here we can indeed observe the convergence as $\beta \rightarrow 0$. This implies that it effectively approximates the scale function and the solution for the CGMY case.

%\begin{figure}[htbp]
%\begin{center}
%\begin{minipage}{1.0\textwidth}
%\centering
%\begin{tabular}{cc}
%(a) $m=100$ & (b) $m=1000$\\
%\includegraphics[scale=0.5]{scale_zero_10}  & \includegraphics[scale=0.5]{scale_zero_100} \\
%Bounds on $W^{(q)}(x)$ when $m=10$ & Bounds on $W^{(q)}(x)$ when $m=100$ \vspace{0.5cm} \\
%\includegraphics[scale=0.5]{derivative_zero_10}  & \includegraphics[scale=0.5]{derivative_zero_100} \\
%Bounds on $W^{(q)'}(x)$ when $m=10$ &  Bounds on $W^{(q)'}(x)$ when $m=100$ \vspace{0.5cm} \\
%\includegraphics[scale=0.5]{u_zero_10}  & \includegraphics[scale=0.5]{u_zero_100} \\
%Bounds on $u(x)$ when $m=10$ & Bounds on $u(x)$ when $m=100$
%\end{tabular}
%\end{minipage}
%\caption{Approximation of the scale function and its derivative when $\sigma = 0$.}
%\label{fig:2}
%\end{center}
%\end{figure}

\appendix

\section{Proofs}

\subsection{Proof of Lemmas \ref{lemma_zeta_wrt_xi} and \ref{lemma_scale_function}} \label{subsection_proof}
By (\ref{dist_x_kappa_2}), it is easy to verify that
\begin{align*}
\E^x \left[ e^{-q \tau_a}1_{\{\tau_a < \infty \}}\right] = \sum_{i =
1}^n \sum_{k=1}^{m_i} A_{i,q}^{(k)} \xi_{i,q} \int_{x-a}^\infty \frac
{(\xi_{i,q} y)^{k-1}} {(k-1)!} e^{-\xi_{i,q} y} \diff y, \quad 0 \leq a < x,
\end{align*}
and hence, because $W^{(q)} \in C^1(0,\infty)$,
\begin{align}
\frac \partial {\partial a}\E^x \left[ e^{-q \tau_a} 1_{\{\tau_a < \infty\}}\right] &= \sum_{i = 1}^n \sum_{k=1}^{m_i} A_{i,q}^{(k)} \xi_{i,q} \frac {(\xi_{i,q} (x-a))^{k-1}} {(k-1)!} e^{-\xi_{i,q} (x-a)}, \quad 0 \leq a < x, \label{eq_derivative_a} \\
\left. \frac \partial {\partial x}\E^x \left[ e^{-q \tau_0} 1_{\{\tau_0 < \infty\}}\right] \right|_{x = 0+} &= - \sum_{i = 1}^n A_{i,q}^{(1)} \xi_{i,q}. \label{eq_derivative_x}
\end{align}
In fact, different representations of (\ref{eq_derivative_a}) and (\ref{eq_derivative_x}) can be pursued. By Theorem 8.1 of \cite{Kyprianou_2006} and (\ref{w_phi}),
 \begin{align*}
 \E^x \left[ e^{-q \tau_a} 1_{\{ \tau_a < \infty \}}\right] = Z^{(q)} (x-a) - \frac q {\zeta_q} W^{(q)}(x-a) = 1 + q \int_0^{x-a} W^{(q)} (y) \diff y - \frac q {\zeta_q} e^{\zeta_q (x-a)}W_{\zeta_q}(x-a)
\end{align*}
for every $0 \leq a < x$.
Its derivative with respect to $a$ becomes
\begin{align}
\begin{split}
\frac \partial {\partial a} \E^x \left[ e^{-q \tau_a} 1_{\{ \tau_a < \infty \}}\right]  &= - q W^{(q)} (x-a) + q e^{\zeta_q (x-a)} W_{\zeta_q}(x-a)  + \frac q {\zeta_q}  {e^{\zeta_q (x-a)}} W_{\zeta_q}'(x-a)    \\ &= \frac q {\zeta_q}  {e^{\zeta_q (x-a)}} W_{\zeta_q}'(x-a).
\end{split} \label{eqn_derivative_x}
\end{align}
In particular, when $a=0$, the derivative with respect to $x$ and its limit as $x \rightarrow 0$ are
\begin{align}
\frac {\partial} {\partial x} \E^x \left[ e^{-q \tau_0} 1_{\{ \tau_0 < \infty \}}\right] = -\frac q {\zeta_q} \left[ -\zeta_q W^{(q)} (x) + W^{(q)'} (x) \right]  \xrightarrow{x \downarrow 0+} - \frac q {\zeta_q} \left[ -\zeta_q W^{(q)} (0) + W^{(q)'} (0+) \right] = - \frac {q} {\zeta_q} \theta. \label{eqn_derivative_0}
\end{align}
By matching (\ref{eq_derivative_x}) and (\ref{eqn_derivative_0}),
Lemma \ref{lemma_zeta_wrt_xi} is immediate.

For the proof of Lemma \ref{lemma_scale_function}, by matching
(\ref{eq_derivative_a}) and (\ref{eqn_derivative_x}) and using Lemma
\ref{lemma_zeta_wrt_xi}, we have

\begin{align*}
W_{\zeta_q}'(y) &=  \frac \theta {\varrho_q} \sum_{i = 1}^n \sum_{k=1}^{m_i} A_{i,q}^{(k)} \xi_{i,q} \frac {(\xi_{i,q} y)^{k-1}} {(k-1)!} e^{-(\zeta_q + \xi_{i,q}) y} \\
&=  \frac \theta {\varrho_q} \sum_{i = 1}^n \sum_{k=1}^{m_i} A_{i,q}^{(k)} (\zeta_q + \xi_{i,q}) \left( \frac {\xi_{i,q}} {\zeta_q + \xi_{i,q}} \right)^{k} \frac {((\zeta_q + \xi_{i,q})y)^{k-1}} {(k-1)!}   e^{-(\zeta_q + \xi_{i,q})y}, \quad y \geq 0.
\end{align*}

Integrating the above and changing variables, we have
\begin{align*}
W_{\zeta_q}(x) - W_{\zeta_q}(0) =\frac \theta {\varrho_q} \sum_{i = 1}^n \sum_{k=1}^{m_i} A_{i,q}^{(k)} \left( \frac {\xi_{i,q}} {\zeta_q + \xi_{i,q}} \right)^{k} \frac 1 {(k-1)!} \int_0^{(\zeta_q + \xi_{i,q})x} {z^{k-1}} e^{-z} \diff z, \quad x \geq 0.
\end{align*}
Lemma \ref{lemma_scale_function} is now immediate because the integral part is a lower incomplete gamma function.

\subsection{Proof of Proposition \ref{proposition_bound_w}}
%We have
%\begin{align*}
%W_{\zeta_q}(x) - W_{\zeta_q}(0) =  \sum_{i=1}^\infty C_{k,q} \left[ 1- e^{-(\zeta_q+\xi_{k,q}) x} \right] = \varrho - \sum_{k=1}^\infty C_{k,q} e^{-(\zeta_q+\xi_{k,q}) x}, \quad x \geq 0.
%\end{align*}
Notice, for every $m \geq 1$, that
\begin{align} \label{bounds_c}
\begin{split}
&0 \leq \sum_{k=1}^m \left( C_{k,q} - C_{k,q}^{(m)} \right) \leq \kappa_q - \sum_{k=1}^m  C_{k,q}^{(m)}  = \delta_m, \\
&0 \leq \sum_{k = m+1}^\infty C_{k,q} =  \kappa_q -\sum_{k =1}^m C_{k,q} \leq \kappa_q -\sum_{k =1}^m C_{k,q}^{(m)} = \delta_m,
\end{split}
\end{align}
and hence by (\ref{bounds_c})
\begin{multline*}
0 \leq \overline{W}_{\zeta_q}^{(m)}(x) -  W_{\zeta_q}(x) = \sum_{k=1}^m \left( C_{k,q}-C_{k,q}^{(m)} \right) e^{-(\zeta_q+\xi_{i,q}) x} + \sum_{k=m+1}^\infty C_{k,q} e^{-(\zeta_q+\xi_{k,q}) x} \\
\leq e^{-\zeta_q x} \sum_{k=1}^m \left( C_{k,q}-C_{k,q}^{(m)} \right)  + e^{-(\zeta_q+\xi_{{m+1},q}) x}\sum_{k=m+1}^\infty C_{k,q}   \leq \delta_m \left[ e^{-\zeta_q x} +  e^{-(\zeta_q+\xi_{m+1,q}) x} \right].
\end{multline*}
Therefore we have the bounds for $W_{\zeta_q}$ in (\ref{bounds_scale}). The bounds for $W^{(q)}$ are immediate by multiplying $e^{\zeta_q x}$.  Finally, the convergence results hold because $\delta_m \rightarrow 0$ and $e^{-\zeta_q x} +  e^{-(\zeta_q+\xi_{m+1,q}) x}$ and $1 +  e^{-\xi_{m+1,q} x}$ are bounded uniformly in $x \geq 0$.

\subsection{Proof of Proposition \ref{proposition_bounds_derivative}}
The lower bound is immediate by the fact that $0 \leq C_{i,q}^{(m)} \leq C_{i,q}$.  For every $x \geq 0$, we have by (\ref{bounds_c})
\begin{align*}
W^{(q)'}(x) &= (\psi'(\zeta_q))^{-1} \zeta_q e^{\zeta_q x} + \sum_{i=1}^\infty C_{i,q}\xi_{i,q}e^{-\xi_{i,q} x} \\
&= \underline{w}^{(m)}(x) + \sum_{i=1}^{m} (C_{i,q}-C_{i,q}^{(m)}) \xi_{i,q}e^{-\xi_{i,q} x}  + \sum_{i=m+1}^\infty C_{i,q}\xi_{i,q}e^{-\xi_{i,q} x} \\
&\leq \underline{w}^{(m)}(x) + \max_{1 \leq k \leq m} (\xi_{k,q}e^{-\xi_{k,q} x})  \sum_{i=1}^{m} (C_{i,q}-C_{i,q}^{(m)}) + \max_{k \geq m+1} (\xi_{k,q}e^{-\xi_{k,q} x})  \sum_{i=m+1}^\infty C_{i,q} \\
&\leq \underline{w}^{(m)}(x) + \left[ \max_{1 \leq k \leq m} (\xi_{k,q}e^{-\xi_{k,q} x})  + \max_{k \geq m+1} (\xi_{k,q}e^{-\xi_{k,q} x})  \right] \delta_m = \overline{w}^{(m)}(x),
\end{align*}
which shows the first claim.  For the second claim, notice that for the given $x_0 > 0$,
\begin{align*}
0 \leq \overline{w}^{(m)}(x) - \underline{w}^{(m)}(x) \leq\left[ \max_{1 \leq k \leq m} (\xi_{k,q}e^{-\xi_{k,q} x_0})  + \max_{k \geq m+1} (\xi_{k,q}e^{-\xi_{k,q} x_0})  \right] \delta_m
\end{align*}
uniformly on $[x_0,\infty)$. Because
$\delta_m$ vanishes as $m \rightarrow 0$ and $\sup_{\lambda \geq 0} \lambda e^{-\lambda x} < \infty$, the convergence is immediate.

\subsection{Proof of Corollary \ref{corollary_z}}
We have
\begin{align*}
W^{(q)'}(x) &= \underline{w}^{(m)}(x) + \sum_{i=1}^m (C_{i,q}-C_{i,q}^{(m)}) \xi_{i,q}e^{-\xi_{i,q} x}  + \sum_{i=m+1}^\infty C_{i,q}\xi_{i,q}e^{-\xi_{i,q} x} \\
&\leq \underline{w}^{(m)}(x) + \max_{1 \leq k \leq m} (\xi_{k,q}e^{-\xi_{k,q} x}) \sum_{i=1}^m (C_{i,q}-C_{i,q}^{(m)})   + \frac {\zeta_q} q e^{-\xi_{m+1,q} x} \sum_{i=m+1}^\infty A_{i,q} \xi_{i,q} \\
&\leq \underline{w}^{(m)}(x) + \max_{1 \leq k \leq m} (\xi_{k,q}e^{-\xi_{k,q} x}) \delta_m + \epsilon_m e^{-\xi_{m+1,q} x}
\end{align*}
where the last inequality holds because
\begin{align*}
\frac {\zeta_q} q \sum_{k = m+1}^\infty  \xi_{i,q} A_{i,q} =  \theta - \frac {\zeta_q} q\sum_{k =1}^m \xi_{i,q} A_{i,q} \leq \theta - \frac {\zeta_q} q\sum_{k =1}^m \xi_{i,q} A_{i,q}^{(m)} = \epsilon_m, \quad m \geq 1.
\end{align*}
This together with Proposition \ref{proposition_bounds_derivative} shows the claim.

\bibliographystyle{apalike}

\bibliography{PhaseTypebib}
\end{document}